\documentclass{aa}

\usepackage{graphicx}
\usepackage{lipsum}
\usepackage{natbib}
\usepackage{amsmath}
\usepackage{multicol}
\usepackage{stfloats}
\usepackage{color}
\usepackage{yfonts}
\usepackage{ulem}
\usepackage{soul} 
\usepackage{hyperref} 
\hypersetup{
    colorlinks=true,
    linkcolor=blue,
    citecolor=teal,
    filecolor=magenta,      
    urlcolor=blue,
    pdfpagemode=FullScreen,
    }

\usepackage{lettrine}

\usepackage{xcolor}

\newcommand{\thalf}{$t_{1/2}$\,}
\newcommand{\teff}{$T_{\rm eff}$\,}
\newcommand{\flatwrm}{\texttt{flatwrm2}\,}
\newcommand{\kepler}{\textit{Kepler}}
\newcommand{\ephemd}{\texttt{ephemd}\,}

\def\addtodot#1.#2\relax{#1\rlap{.}^{\dotadd}#2}

\newcommand{\dotem}[1]{\def\dotadd{m}\addtodot#1\relax}

\usepackage{txfonts}
\begin{document}

   \title{Stellar flare morphology with TESS across the main sequence}


\author{B.~Seli
          \inst{1,2,3}
        \and
          K.~Vida
          \inst{1,2}
        \and
          K.~Ol\'ah
          \inst{1,2}
        \and
         A.~Görgei
          \inst{1,2,3}
        \and
         Sz.~Soós
          \inst{3,4}
        \and
         A.~Pál
          \inst{1,2,5}
        \and
         L.~Kriskovics
          \inst{1,2}
        \and
         Zs.~K\H{o}v\'ari 
          \inst{1,2}
  }

   \institute{Konkoly Observatory, HUN-REN Research Centre for Astronomy and Earth Sciences, Konkoly Thege Mikl\'os \'ut 15-17., H-1121, Budapest, Hungary\\
   \email{seli.balint@csfk.org}
        \and
            HUN-REN CSFK, MTA Centre of Excellence, Budapest, Konkoly Thege Mikl\'os út 15-17., H-1121, Budapest, Hungary
        \and
            E\"otv\"os University, Department of Astronomy, Pf. 32, H-1518, Budapest, Hungary
        \and
            Gyula Bay Zolt\'an Solar Observatory (GSO), Hungarian Solar Physics Foundation (HSPF), Pet\H{o}fi t\'er 3, H-5700 Gyula, Hungary
        \and   
            Eötvös Loránd University, Institute of Physics and Astronomy, H-1117 Budapest, Hungary
            }

   \date{Received October 4, 2024; accepted December 16, 2024}

 
  \abstract
   {Stellar flares are abundant in space photometric light curves. As they are now available in large enough numbers, the statistical study of their overall temporal morphology is timely.}
   {We use light curves from the Transiting Exoplanet Survey Satellite (TESS) to study the shapes of stellar flares beyond a simple parameterization by duration and amplitude, and reveal possible connections to astrophysical parameters.}
   {We retrain and use the \flatwrm long-short term memory neural network to find stellar flares in 2-min cadence TESS light curves from the first five years of the mission (sectors 1--69). We scale these flares to a comparable standard shape, and use principal component analysis to describe their temporal morphology in a concise way. We investigate how the flare shapes change along the main sequence, and test whether individual flares hold any information about their host stars. We also apply similar techniques to solar flares, using extreme ultraviolet irradiation time series.}
   {Our final catalog contains $\sim 120,000$ flares on $\sim 14,000$ stars. Due to the strict filtering and the final manual vetting, this sample contains virtually no false positives, although at the expense of reduced completeness. Using this flare catalog, we detect a dependence of the average flare shape on the spectral type. These changes are not apparent for individual flares, only when averaging thousands of events. We find no strong clustering in the flare shape space. We create new analytical flare templates for different types of stars, present a technique to sample realistic flares, and a method to locate flares with similar shapes. The flare catalog, along with the extracted flare shapes, and the data used to train \flatwrm are publicly available.}
   {}

   \keywords{stars: activity --
            stars: flare --
            stars: statistics --
            Sun: flares
               }

   \maketitle
%

\section{Introduction}

The advent of space photometry enabled detailed statistical studies of stellar flares in volumes never seen before \citep[e.g.,][]{2014ApJ...797..121H, 2016ApJ...829...23D, 2017ApJ...849...36Y, 2018ApJ...868....3R, 2019ApJS..241...29Y, 2020AJ....159...60G, 2021A&A...647A..62O, 2022ApJ...925L...9F}. The \kepler{} space telescope \citep{2010Sci...327..977B} observed the same field over four years, providing accurate flare statistics for thousands of stars. The Transiting Exoplanet Survey Satellite \citep[TESS,][]{2014SPIE.9143E..20R} observes the whole sky in 27 days-long sectors, providing shorter light curves of even more objects. While the \kepler{} observatory was designed to survey a portion of the sky to discover Earth-like exoplanets, and its targets were mainly solar-type stars, TESS observes almost the whole sky and has more late-type stars as targets, including more flaring M dwarfs.

Stellar flares are the most easily observable manifestations of magnetic activity \citep{1989SoPh..121..299P, 2024LRSP...21....1K}. They appear on light curves as sudden bursts, lasting for minutes or hours. Using photometric data in a single filter, it is possible to determine the time of the flare peak, its amplitude, duration, and the energy released in the given filter.

Most flare studies focus on flaring rate and energy distribution on different kinds of active stars \citep[see, e.g.,][]{2014ApJ...792...67C, 2019ApJS..241...29Y, 2020AJ....159...60G, 2023A&A...669A..15Y, 2022ApJ...925L...9F,2024AJ....168...60F, 2024MNRAS.527.8290P}, looking for a dependence on spectral type, age, rotational period, and other stellar parameters. Other applications of basic flare properties include the search for changes in flare rate \citep{2022ApJ...941..193C}, search for periodicity in flaring times \citep{2021ApJ...920...42H}, or the study of waiting time and rotational phase distributions \citep{2014ApJ...797..121H,2020MNRAS.494.3596D}.

When the emphasis is on studying the temporal morphology of flares, higher cadence observations are necessary, where the flare events are resolved in time. Using space photometry, the following options are available: 1 and 30-min cadence for Kepler, 20-s and 2-min for TESS short cadence mode (for pre-selected targets), and 200-s, 10-min and 30-min for TESS full frame images. While \kepler{} and TESS are the most popular space-based options, there are flare-related studies using MOST \citep{2012PASP..124..545H, 2016ApJ...829L..31D} and CHEOPS \citep{2024A&A...686A.239B} data.

One of the most influential studies about stellar flare profiles was presented by \cite{davenport_template}. Using 1-min cadence \kepler{} light curve of the M4 dwarf GJ~1243, \cite{davenport_template} created a flare profile template, combining a polynomial rise phase with a double exponential decay phase. This template was extensively used to model stellar flares observed with different instruments \citep[e.g.,][]{2022A&A...668A.111M, 2022ApJ...936...17H, 2022ApJ...935..104M, 2022MNRAS.513.2615M, 2023MNRAS.519.3564J}. An updated flare model of the same star was introduced by \cite{2022AJ....164...17M}, using the convolution of a Gaussian and a double exponential to create a profile that is differentiable at the peak. A detailed study of flare shapes was carried out by \citet{2022ApJ...935..143P} on a sample of 140,000 TESS flares. They used multiple flare profile models, using the convolution of a Gaussian and an exponential decay, and combining two of these profiles. \cite{flaring_giants} contrasted flares of dwarfs and giants, and found that while the distribution of their durations are different, their profiles are similar, at least with 30-min cadence \kepler{} data.

\cite{2024A&A...686A.239B} comprehensively analyzed 20-s cadence TESS and 3-s cadence CHEOPS light curves, and revealed that a significant fraction of flares are complex with the adequate time resolution. They separated multi-peak flares to study the individual components and also identified possible quasi-periodic pulsations and a pre-flare dip. Based on TESS 20-s data, \cite{2022ApJ...926..204H} showed that a large fraction of flares have a substructure during the rising phase, and short-period quasi-periodic pulsations are quite common. They also found that a significant fraction of flares have a gradual Gaussian peak following the primary impulsive peak.

The previous studies presented results on large ensembles of flares, but in some cases, individual events are also analyzed, for example, in the case of quasi-periodic pulsations \citep{2020ApJ...905...70P, 2022MNRAS.514.5178D}. To study the variability of stellar flares on a timescale of seconds, ground-based measurements are available, with much smaller sample sizes (see e.g., \mbox{\citealt{2016ApJ...820...95K, 2022PASJ...74.1069A}}). Focusing on a limited number of targets also makes it possible to obtain multi-band or spectral data simultaneously (see e.g., \mbox{\citealt{2023JAVSO..51...14B}, \citealt{2013ApJS..207...15K}}).

Flare profiles can also be analyzed on the Sun, where more detailed observations are possible. \cite{2021MNRAS.502.3922K} studied the temporal morphology of solar flares, with Sun-as-a-star flux measurements from the \textit{Solar Dynamics Observatory}'s \textit{Atmospheric Imaging Assembly} (SDO/AIA). \cite{2017SoPh..292...77G} studied solar X-ray flare light curves, and defined the flare profile as the convolution of a Gaussian heating and an exponential decay, thus having a smoothly varying profile that fits high cadence solar observations well.

To explain the diversity of flare shapes, a few theoretical models were proposed. \citet{2003A&A...399..647T} put foward a geometric model. They treat a flare as a short impulsive event that heats the base of the associated magnetic structure in the photosphere, which then radiates more gradually, giving rise to typical "peak-bump" shapes \citep{2022ApJ...926..204H}. Then, depending on the position of the footpoint (or \textit{echo}) on the stellar disk, different temporal morphologies may be observed. \cite{2023ApJ...959...54Y} also modelled "peak-bump" flares, using one-dimensional hydrodynamic loop simulations, and found that radiating plasma from the loop can contribute to the secondary peak in the optical.

In this work, we intend to gain further empirical insights about the temporal morphology of stellar flares. We use TESS light curves to compile a large, homogeneous and pure sample of stellar flares. After scaling these flares to a standard shape, we apply dimensionality reduction techniques to summarize the information carried by the morphology of these flares, beyond the simple parameterization by duration and amplitude. This representation of the flare shape is parameter-free, unlike models involving multiple polynomial, Gaussian, etc. components. We then try to find regularities in the flare shapes, including clusters of different shapes or correlations with astrophysical parameters. Finally, we apply similar techniques for solar flares observed by the \textit{Extreme Ultraviolet Variability Experiment} (EVE) instrument of SDO, to look for obvious differences between flares produced under different conditions.

\section{Data and methods}

\subsection{TESS}
To search for stellar flares, we used 2-min cadence \textit{Pre-search Data Conditioning Simple Aperture Photometry} (PDCSAP) light curves provided by the SPOC pipeline \citep{10.1117/12.2233418} from the first five years of the TESS mission, up to sector 69.

Since the goal of this work is not to compile a complete catalog of TESS flares, but to study their average shapes, we excluded the noisier light curves from the start. To this end, we smoothed each available TESS light curve with a 31-point (one hour) wide running median filter, and kept it only if the ratio of the standard deviation of the smoothed and original datasets exceeded an empirically derived threshold of 0.4, indicating that astrophysical variation dominates short-timescale random noise:
\begin{equation}
    \sigma_{\mathrm{ratio}} = \frac{\mathrm{STD_{smoothed}}}{\mathrm{STD_{original}}} > 0.4
\end{equation}
This way, based on the manually vetted training set (introduced in the following section), we can exclude $\sim$60\% of the available TESS light curves to speed up the computation, while only losing $\sim$10\% of the flaring stars.

\subsection{Flare detection method}

Several automated tools exist to identify stellar flares in light curves, including the use of convolutional neural networks \citep{2020JOSS....5.2347F, 2022ApJ...935...90T, 2024arXiv240721240J}, Bayesian odds ratio \citep{2014MNRAS.445.2268P}, differencing \citep{2022ApJ...935..102B}, RANdom SAmple Consensus \citep[RANSAC,][]{2018A&A...616A.163V}, multi-algorithm voting \citep{2024AJ....168..234L}, and hidden Markov models \citep{2024arXiv240413145E, 2024MNRAS.534.2142Z}.

In this work, we used \flatwrm \citep{flatwrm2}, a long short-term memory (LSTM) neural network originally developed to find flares in \kepler{} light curves, with the emphasis on low astrophysical false positive rate from known variable stars, such as RR Lyrae or eclipsing binaries.

\begin{figure}[ht]
\includegraphics[width=\columnwidth]{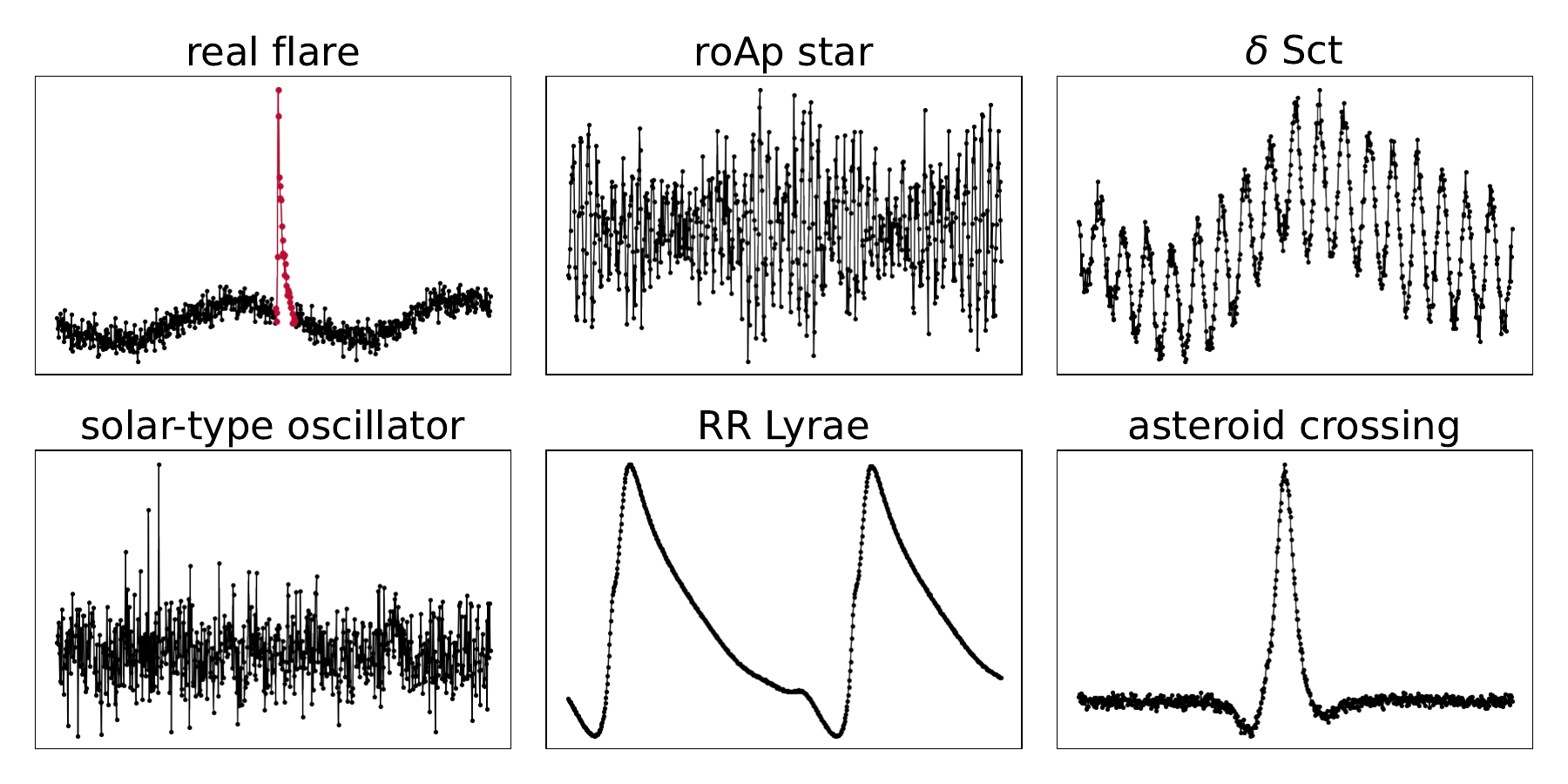}
\caption{Example light curves from the training set. The upper left panel shows a real flare, the others are false positives. All panels show one-day-long segments.}
\label{fig:training_set_examples}
\end{figure}

We retrained \flatwrm specifically to TESS 2-min cadence data, with the original architecture and an augmented training set. Apart from the original training set, we added 4631 TESS light curves from sectors 1--69 with flares identified manually. These include random stars, stars that are expected to flare, and also typical false positives. We collected flaring candidates from \cite{2020AJ....159...60G}, previous runs of \flatwrm, and also from the following TESS Guest Observer proposals: G011266, G04039, G04139, G05105, G03227, G04051, G04234. For the false positives, we added a few hundred stars from these sources: rapidly oscillating Ap stars from \cite{roap_tess}, $\delta$ Scuti hybrids from \cite{marek_tess}, solar oscillators from \cite{solar_oscillators_tess}, RR Lyrae stars from the TESS G03169, G04106 and G04184 proposals, and stars with solar system asteroids moving through the aperture \citep[as identified by the \ephemd{} tool, see][]{2020ApJS..247...26P}. Figure~\ref{fig:training_set_examples} shows a few examples of the "astrophysical noise" set. We inspected each light curve and flagged flaring points using a box selection tool, resulting in an array of ones and zeros for flaring and non-flaring points. Most of the manually vetted light curves included no flares, so to balance the training set, we excluded $\sim$2/3 of the non-flaring stars. The final set of 4631 light curves includes the following: 50\% flaring, 34\% non-flaring with $\sigma_{\mathrm{ratio}}>0.4$, 2\% non-flaring with $\sigma_{\mathrm{ratio}}<0.4$ and 14\% false positives.

We trained \flatwrm on this new training set using $k$-fold cross-validation, following the same procedures as \cite{flatwrm2}. We used this retrained version of \flatwrm to find flares in the first 69 sectors of TESS 2-min data. In the following sections, we describe the post-processing steps and the content of the final flare catalog.

To facilitate future data-driven efforts for stellar flare detection, we make the manually flagged light curves publicly available on Zenodo\footnote{\url{https://zenodo.org/records/14179313}}, as a series of time, flux, and 0/1 flags for each light curve. The fully trained model and the weight file is also available on Github\footnote{\url{https://github.com/vidakris/flatwrm2}}.

\subsection{Post-processing of the \flatwrm results}
\label{sect:post-proc}

\begin{figure}[ht]
\includegraphics[width=\columnwidth]{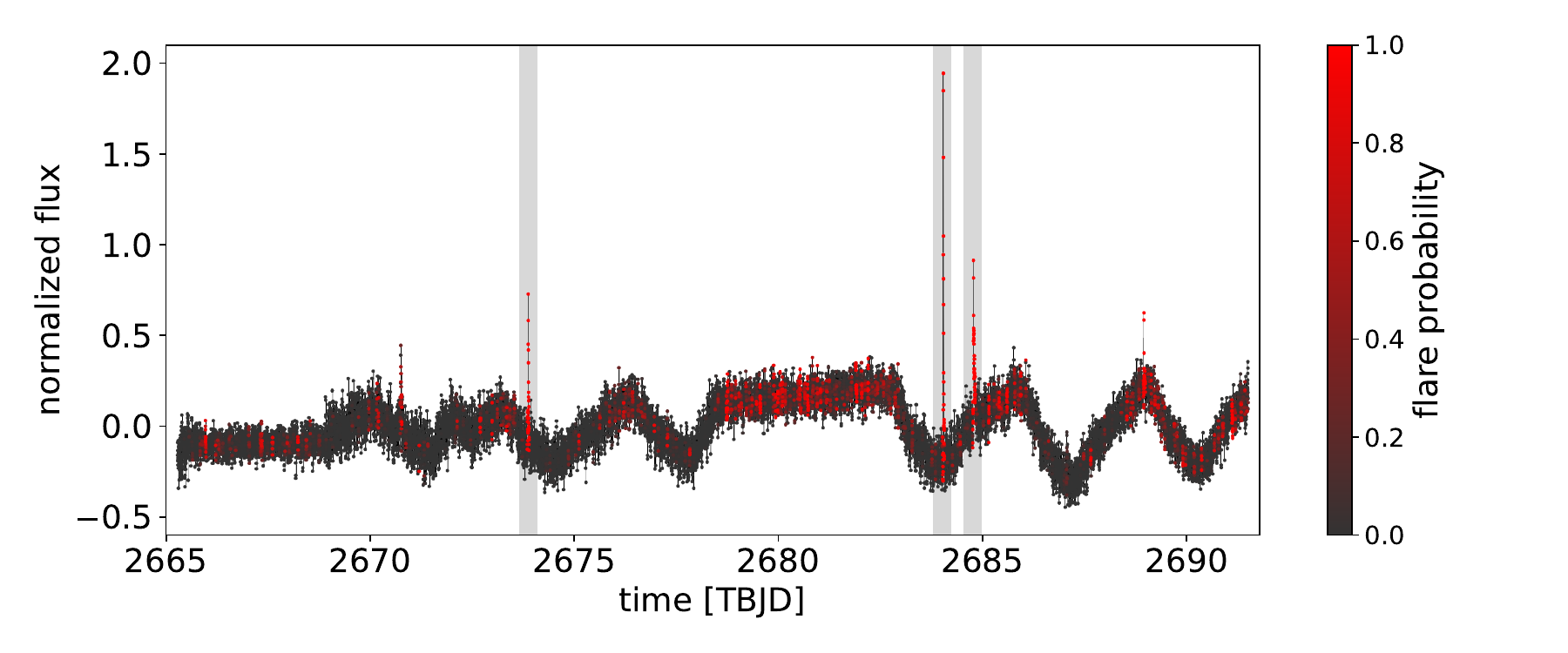}
\caption{Example light curve color coded with the \flatwrm prediction. Gray lines show the positions of the validated flares from the final catalog.}
\label{fig:flatwrm2_example}
\end{figure}

The raw output of \flatwrm is a flare probability time series (see Fig.~\ref{fig:flatwrm2_example} for an example). To extract individual flare events from this output, we run the \flatwrm validation step (see \citealt{flatwrm2} for details). Running this validation step on the 444,963 TESS light curves with $\sigma_{\mathrm{ratio}}>0.4$ resulted in 3,103,728 flare candidates.

As this initial candidate list contains many false events, we filtered it based on the $A$ flare amplitude, ED (equivalent duration, measured in days), and S/N parameters available from \flatwrm. After manually inspecting a few hundred candidates, we used the following criteria to remove smaller events and brightenings from other astrophysical sources:

\begin{itemize}
    \item ${\rm S/N} > 5$
    \item $A > 0.001$
    \item $0.001 \cdot A < {\rm ED} < 0.1 \cdot A$
\end{itemize}

To clean the flare catalog further, we employed three more criteria for each candidate: i) the peak must rise above 3 standard deviations from the median of the quiescent light curve, after removing a parabolic trend; ii) there must be no NaN points in the 15\,min vicinity of the peak; iii) there can be no more than one point flagged with the bitmask 6591. This bitmask includes the following quality flags \citep{tess_doc_2020}: Attitude tweak, Safe mode, Coarse point, Earth point, Argabrightening, Desaturation event, Manual exclude, Discontinuity corrected, Straylight and Straylight2. We did not use the following quality flags, as they would sometimes remove real flare peaks, as also noted by \cite{2020AJ....160..219F}: Impulsive outlier, Cosmic ray in collateral data, Cosmic ray in optimal aperture.

As a next step, we extract and scale the flare events in time and flux. Table~\ref{table:sample_size} shows the number of flare candidates after each processing step.

\begin{table}
\caption{The sample size after each processing step}
\label{table:sample_size}
\centering
\small
\begin{tabular}{lr}
\hline \hline
TESS 2-min light curves (up to sector 69): & 1,258,154 \\
TESS 2-min light curves with $\sigma_{\mathrm{ratio}}>0.4$: & 444,963 \\
Flares found by \flatwrm: & 3,103,728 \\
Flares filtered from \flatwrm: & 294,357 \\
Flares after post-processing\\ \hspace{0.1cm}(peak, no NaNs, no bad quality flags): & 212,228 \\
Flares successfully extracted: & 148,887 \\
Correctly extracted, manually vetted flares: & \textbf{121,895} \\
(Incorrectly extracted and blacklisted flares: & 14,649) \\
\hline
\end{tabular}
\end{table}

\subsection{Flare extraction from the light curves}
\label{sect:extraction}

Once we know where the flares are, we need to extract them from the quiescent baseline variation, along with some basic parameters, i.e., amplitude, some measure of length, and ED. To model the baseline variation, local polynomial fits, linear fits \citep{davenport_template} and Gaussian processes are commonly used \citep[e.g.,][]{2022AJ....164...17M, 2022AJ....163..147G}.

\begin{figure}[ht]
\includegraphics[width=\columnwidth]{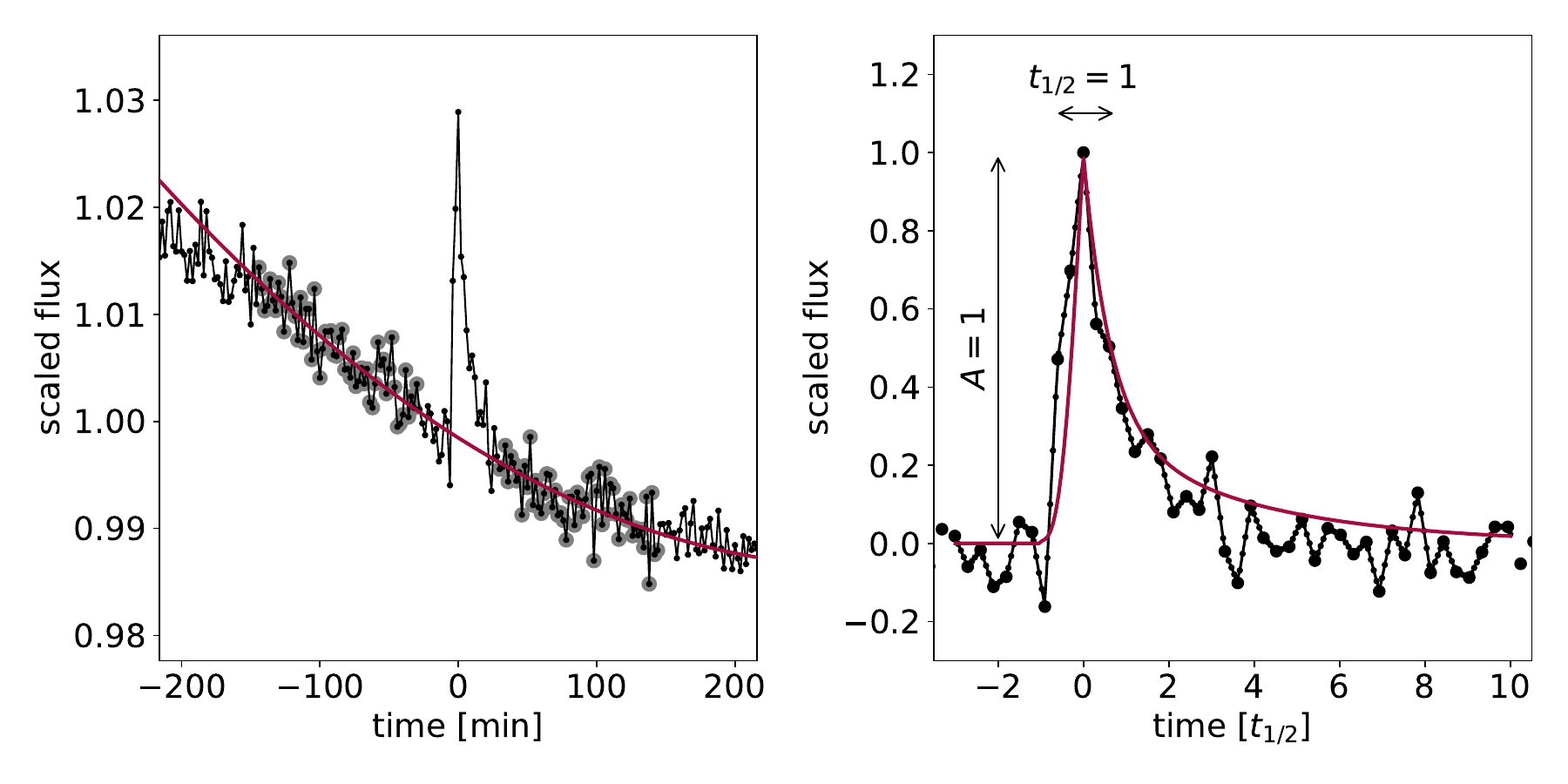}
\caption{An illustrative example of the extraction of a scaled flare shape. \textit{Left:} Gray shows the points used for the baseline fit, red line shows the fitted polynomial. \textit{Right:} The red line shows the flare template used for the time scaling. The large black dots are from the original light curve, the small black dots are the interpolated points.}
\label{fig:extraction}
\end{figure}

To start, we centered the flare peak time to zero. As there might be a slight offset from the peak time from \flatwrm, we repositioned it to the maximum in the 30\,min vicinity. Then, we cut a $\pm 0.1$\,days segment around it, and masked out the duration of the flare ($t_{\rm peak} \pm 30$\,min). We clipped a remaining segment with $2 \sigma$ to remove any residual variation by the flare. Then, we fitted a polynomial to this quiescent light curve with a degree between 0 and 4 as favored by the minimal Bayesian information criterion (BIC, \citealt{2007MNRAS.377L..74L}), defined as:
\begin{equation}
    \mathrm{BIC} = n \ln{\frac{\sum_i \left( y_i - \hat{y_i} \right)^2}{n}} + k \ln{n},
\end{equation}
where $n$ is the number of data points, $k$ is the degree of freedom (polynomial degree plus one), $y_i$ and $\hat{y_i}$ are the measured and modeled points.
After removing this polynomial trend, the flare was fitted with the single-peaked flare template of \cite{davenport_template} to estimate the \thalf time scale of the flare, which is the full width at half maximum of the template. This template is given in the following form, after transforming the measured time $t$ to $t' = \frac{t - t_\mathrm{peak}}{t_{1/2}}$:
\begin{equation}
\small
    \begin{cases} 0 &\mbox{if } t' < -1 \\
A \cdot (1 + 1.941 t' - 0.175 t'^2 - 2.246 t'^3 - 1.125 t'^4) & \mbox{if } -1 \leq t' < 0 \\
A \cdot (0.689 \cdot e^{-1.6 t'} + 0.303 \cdot e^{-0.2783 t'}) & \mbox{if } 0 \leq t' 
\end{cases}
\end{equation}
After centering the flare peak time to zero and scaling time with the fitted \thalf, we linearly interpolated the segment to a grid of 200 points between $-3$ and $10 $  \thalf. As a final detrending step, we removed a linear fit from the interpolated segment, fitted before $-t_{1/2}$ and after $8t_{1/2}$. The amplitude of the flare was scaled to unity (see Fig.~\ref{fig:extraction}). Similar steps were followed by \cite{flaring_giants}. We calculated the ED from the baseline-removed light curve, using the trapezoidal rule for integration. Also, we calculated the signal-to-noise ratio (S/N) of the flare, by dividing the amplitude by a local measure of scatter. To calculate the scatter, we differenced the 0.2\,days vicinity of the flare, and took half the difference between the 16th and 84th percentiles ($1\sigma$ in the Gaussian case).

The given flare was discarded if \thalf$<2$\,min or if S/N < 3, removing $\sim 30$\% of the candidates. Finally, we saved the scaled and interpolated flare shape, amplitude, \thalf, ED and S/N. The amplitude and \thalf were also measured by \flatwrm, but we recalculated them here, to match the extraction method.

We note that using a flare template for scaling could cause a bias in the resulting average shape, as it enforces a given shape onto the flare events. For example, using a triangle-shaped template results in "boxy" flares. However, since we are interested in the relative differences, it is not a problem as long as the same template is used for all the events.

\subsection{Manual vetting}
\label{sect:manual_vetting}

After the filtering and extraction steps described in the previous sections, we are left with 148,887 scaled flare profiles. This list is already relatively pure, however, defects can occur during the extraction process, resulting in deformed flare shapes. As we are interested in subtle differences in the flare profiles, we add a final manual vetting step to the analysis. We visually inspected every single extracted flare and classified them into three groups: i) correctly extracted real flare, ii) incorrectly extracted real flare (e.g., with defects in the baseline removal, too many missing points), iii) non-flare (e.g., nova-like flickering). For each candidate, we made this judgment by plotting the 2\,days, 7\,hours and 4\,hours vicinity of the event, and also the final scaled and interpolated profile. We use the correctly extracted real flares in the remainder of the manuscript without any modifications. We make no further attempt to correct the erroneously extracted flares, but we list their $t_{\rm peak}$ in the final catalog. The non-flares are removed immediately from the sample.

A noteworthy case is that of the complex flares. It is still debated whether these events are just by-chance alignments of individual flares, or whether there is a physical connection between them \citep[see, e.g.,][]{2011ApJ...739L..63T}, so we do not remove them from the sample by default. However, as many of them are hard to extract, a large fraction of the complex flares will be missing from the final sample anyway.

As stars on the Hertzsprung--Russell diagram are not equally likely to produce observable flares, certain types of stars -- where one would less expect flaring activity --  warrant extra scrutiny (see \citealt{2021A&A...647A..62O} for a discussion on flaring giants). We selected the following objects based on their Gaia properties, and repeated the classification of their correctly extracted flares:
\begin{itemize}
    \item hot stars: $G_{BP}-G_{RP} < 0.5$
    \item giant stars: $3 \cdot (G_{BP}-G_{RP}) - 1.5 > M_G$
    \item subdwarfs and white dwarfs: $3 \cdot (G_{BP}-G_{RP}) + 3.5 < M_G$
\end{itemize}
This selection included 1433 flares, out of which we discarded 612. After this check, a few flares still remained on these objects. It is still possible that these flares do not originate from the given stars, but are from unresolved companions or other contaminating objects. However, we note that the flare amplitudes are noticeably higher on white dwarfs, which is consistent with their low luminosities producing higher flare contrast. We make no further attempt to validate these flares, as it is not the main focus of this paper, but we encourage interested readers to further examine these objects.

The final occurrence rate of the three categories is 83\% for correctly extracted real flares, 10\% for incorrectly extracted real flares, and 7\% for non-flares.

\subsection{Duplicate flares}

The manual vetting process revealed that the catalog contains duplicate entries. This can happen for two different reasons. The first one is that there are duplicate light curves in TESS 2-min cadence data, due to the large (21'') pixel size of TESS. This way, a bright flaring star can contaminate the neighbouring pixels, causing (almost) the same flares to appear on different stars at the exact same time. Getting rid of these flares is beyond the scope of this paper. Apart from identifying the duplicates, the main challenge is to which star should the flare be attributed. The solution would require the use of pixel-level data, as in the case of \cite{2022ApJ...935...90T} and \cite{2023AJ....165..141H}.

The second reason is that during the extraction procedure (see Sect.~\ref{sect:extraction}), the flare peak was shifted from its initial position from \flatwrm to the light curve maximum in a 30\,min window, to center the peak. This way, some neighbouring flare candidates have been merged. We remove these duplicates from the catalog, totaling 1065 events, out of which 688 were correctly extracted real flares (see Sect.~\ref{sect:manual_vetting}).

\subsection{Blacklisted objects}

One needs to be careful when interested in the flares with the highest amplitude or energy. The flare amplitude (and also the ED) is measured as a flux increase compared to the quiescent level. If the quiescent level is erroneous for some reason, for example, due to defects in the background removal in the TESS photometry pipeline, one can measure extremely high flare amplitudes. This was the case for TIC\,231799463 and TIC\,1801578770, where the flare amplitudes ranged from a few to even a hundred times the quiescent level. We identified these objects on the TESS magnitude vs. light curve noise plot as outliers. We extracted their light curves from the target pixel files with a different pipeline (\texttt{eleanor}, \citealt{2019PASP..131i4502F, 2019ascl.soft05007B}), and got a higher quiescent flux baseline, thus smaller, more realistic flare amplitudes. We removed the measured flare parameters of these two stars from the catalog. We note that there are probably more stars affected by erroneous background removal, but these two were the most extreme.

\subsection{Astrophysical parameters and flare energies}\label{sect:parameters}

We collected the following astrophysical parameters from version 8.2 of the TESS Input Catalog \citep[TIC,][]{2019AJ....158..138S}:  effective temperature ($T_\mathrm{eff}$),  surface gravity ($\log g$), bolometric luminosity ($L_\mathrm{bol}$), Gaia DR2 \citep{2018A&A...616A...1G} $G_{BP}-G_{RP}$ color index, Gaia DR2 $M_G$ absolute $G$ magnitude calculated from the observed $G$ magnitude and the parallax.

To calculate flare energies from EDs (area below the flare on the normalized light curve), we used the same approach as in \citet{flaring_giants}. We collected BT-NextGen model spectra \citep{nextgen} in the $T_\mathrm{eff}$ and $\log g$ range of the sample with solar metallicity. For each star, we selected the closest spectra in the $T_\mathrm{eff}$--$\log g$ grid, integrated over the whole wavelength range with and without convolving it with the TESS response function, so the ratio of the TESS to bolometric luminosity can be calculated. Using $L_\mathrm{bol}$ from TIC, the $L_\mathrm{TESS}$ quiescent luminosity in the TESS band can be calculated. Then, the flare energy in the TESS band is given by:
\begin{equation}
    E_{\rm{TESS}} = L_{\rm{TESS}} \cdot \rm{ED}.
\end{equation}
For stars with no $\log g$ in TIC, we assumed $\log g=4.7$, the sample median. For stars with missing $L_\mathrm{bol}$, we estimated it from the absolute $G$ magnitude. These guesses are only used for the estimation of $L_\mathrm{TESS}$, and not used anywhere else. They affected 13\% of the sample, and are estimated to increase the uncertainty of $L_\mathrm{TESS}$ by $\sim10$\%.

We also tried two different methods for the calculation of $L_\mathrm{TESS}$. One is simply using a black body spectrum with the \teff from TIC, instead of using a BT-NextGen model spectrum. The other method uses the apparent $T$ magnitude and distance of each star to calculate $L_\mathrm{TESS}$ directly. For this, we need the apparent $T$ magnitude and the TESS band luminosity of the Sun. Using a standard solar spectrum and the TESS response function, we get $L_{\rm{TESS},\odot}=1.03 \cdot 10^{33}$\,erg\,s$^{-1}$, and by transforming the solar Gaia magnitudes \citep{2019AJ....158..138S}, we get $T_{\odot}=\dotem{-27.3}$. All three methods agree within a few percent down to $T_{\rm eff} \approx 4000$\,K, below which 20--40\% differences can occur.

\subsection{Dimensionality reduction}
\subsubsection{Weighted principal component analysis}

To summarize the information contained in the scaled flare shapes, and to visualize any trends with astrophysical parameters, we use principal component analysis (PCA, \citealt{pca_pearson}, or for recent applications see e.g., \citealt{2018ApJ...857...55H, 2021MNRAS.502.5762C, 2022A&A...659A...3S}). PCA is a linear method able to reduce the dimensionality of the interpolated flare shapes from 200 to a few. It defines a new basis with vectors (principal components) pointing in the direction of the highest variance. Using this new basis, a large fraction of the sample variance can be recovered by using only the first few principal components. This way, it is possible to describe the shape of the flares in a model-free way, without assuming any analytical functional form.

Since the scaled shapes of longer flares are better sampled than shorter ones, they are more valuable in the analysis. Similarly, flares with higher S/N are also more important. To account for this, we employ weighted principal component analysis \citep[WPCA,][]{wpca}\footnote{Implemented in the \texttt{wpca} python package: \url{https://github.com/jakevdp/wpca}}. We use the following weighting factor for each flare, with \thalf measured in days:
\begin{equation}
    W_i = \log_{10} \left( t_{1/2, i} \cdot {\rm S/N}_i \right)
\end{equation}
This is only slightly different than using uniform weights, the 1st, 50th and 99th percentiles are 0.38, 0.96 and 2.24, respectively. This means that the weights span roughly one order of magnitude. We note that WPCA could weigh each point of each flare differently, but we use uniform weights across the points of individual flares.

\begin{figure*}[ht]
\includegraphics[width=2\columnwidth]{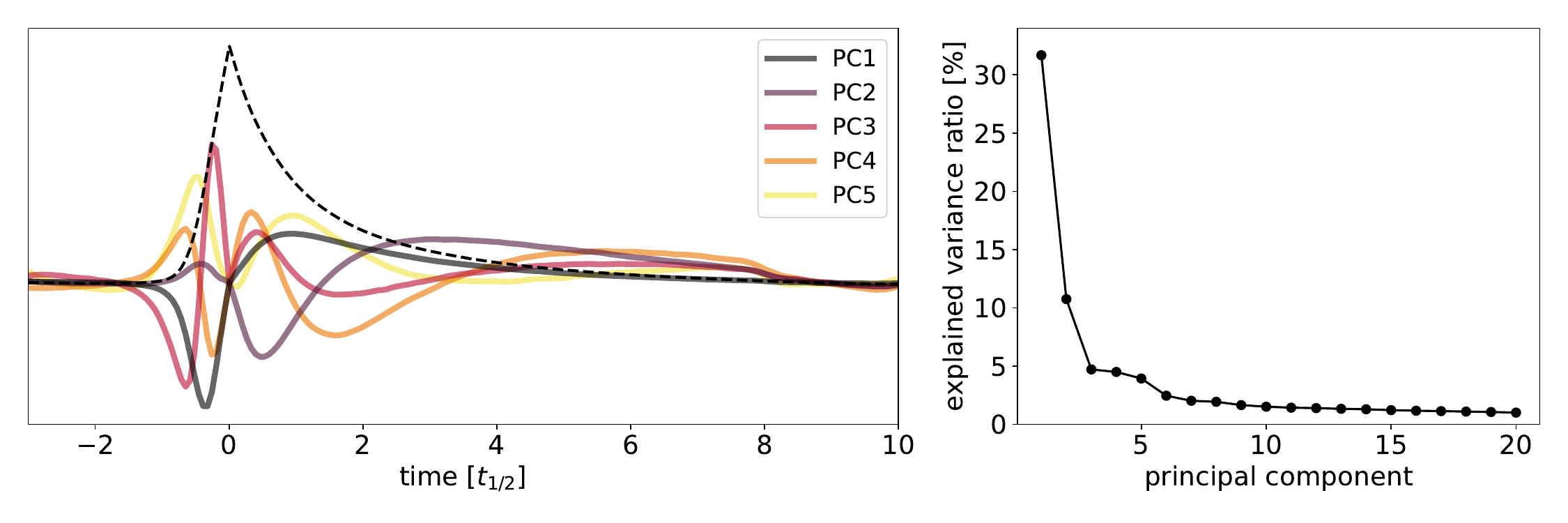}
\includegraphics[width=2\columnwidth]{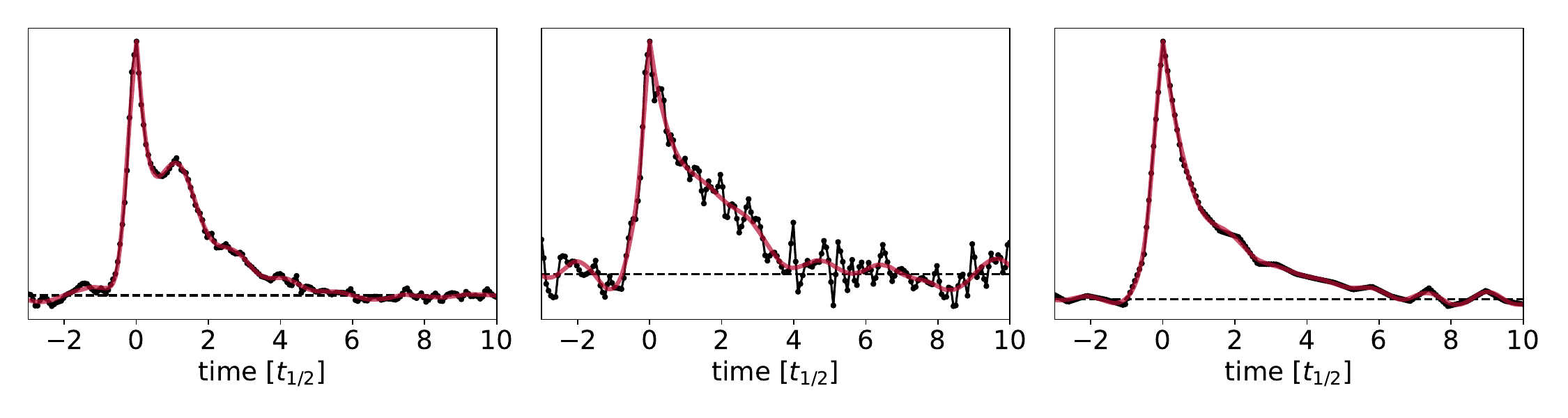}
\caption{The weighted PCA basis. \textit{Upper left:} The first 5 principal components, with a dashed line denoting the average flare profile. \textit{Upper right:} The ratio of the sample variance that a given PC can recover. A single feature from the original 200-dimensional dataset would amount to 0.5\%. \textit{Lower panels:} Example light curves with the PCA reconstruction using 20 PCs.}
\label{fig:pca_components}
\end{figure*}

Figure~\ref{fig:pca_components} shows the first 5 principal components (PCs), and the importance of each PC to explain the sample variance (upper right panel). The fact that all PCs are flat before $-$\thalf and after 8\thalf is an artifact of the extraction, as the final detrending line was fitted to those regions (see Sect.~\ref{sect:extraction}). It can be seen that even the first 3 PCs can recover 47\% of the variance and that there is an "elbow point" at 3 PCs, after which new PCs contain less information. This suggests that the flare profiles can broadly be described by only a few parameters. There is possibly an additional elbow point at 6 PCs, and as we will see later, it might be linked to the stellar \teff. The lower panels of Fig.~\ref{fig:pca_components} show the PCA reconstruction of a few flare events. For the low S/N cases, PCA also acts as a simple denoising, keeping only the more important features. Throughout the paper, we use 5 PCs for visualization purposes (56\% explained variance), and 20 PCs for calculations (77\% explained variance, e.g., for the astrophysical parameter estimation).

\subsubsection{Uniform Manifold Approximation and Projection}

As PCA is a linear algorithm, it might struggle to find inherently nonlinear relationships (e.g., a Swiss roll shape in 3D). A powerful nonlinear dimensionality reduction algorithm is Uniform Manifold Approximation and Projection
\citep[UMAP,][]{2018arXiv180203426M}\footnote{Implemented in the \texttt{umap} python package: \url{https://github.com/lmcinnes/umap}}, which is fast and scalable to larger datasets. Other popular options include Isomap, t-SNE and autoencoders, see \cite{2019arXiv190407248B} for an overview. Nonlinear dimensionality deduction can provide a more compact, lower dimensional representation of the dataset than PCA, although they are less robust, more sensitive to noise, and depend strongly on the random seed. We use UMAP only for visualization purposes.

\section{TESS results}

\subsection{The final flare catalog}

Following the filtering steps on the \flatwrm results, our final sample includes 121,895 correctly extracted flares on 14,408 stars (see Table~\ref{table:sample_size} for the sample size after each step). With the final manual vetting step, the catalog has virtually no false positives, at the expense of relatively low completeness. This makes it suitable for our purposes, but it might not be the optimal choice for many other studies (e.g., calculating flare rates, creating flare frequency distributions), as weak flares with low S/N ratio, or even larger flares with incorrect extraction are removed.

The stellar astrophysical parameters are presented in Table~\ref{table:tess_stars}, and the flare catalog is presented in Table~\ref{table:tess_flares}. These tables also include the incorrectly extracted flares identified during the manual vetting, but in those cases, we omit the \thalf, amplitude, ED and $E_{\rm TESS}$ parameters, as those are likely erroneous. The scaled flare shapes are also available online.

\begin{table}
\caption{Astrophysical parameters of the flaring stars. The color and absolute magnitude are from Gaia DR2, the \teff and $\log g$ values are from TICv8.2, and the quiescent stellar luminosity is calculated as described in Sect.~\ref{sect:parameters}. The last column indicates the number of sectors the target was observed with 2-min cadence up to sector 69. The full table is available online.}
\label{table:tess_stars}
\centering
\tiny
\tabcolsep=0.11cm
\begin{tabular}{cccccccc}
\hline \hline
TIC & $G_{\rm BP}-G_{\rm RP}$ & $M_G$ & \teff & $\log g$ & $L_{\rm TESS}$ & \#flares & \#sectors\\
 & [mag] & [mag] & [K] & & [erg s$^{-1}$]\\
\hline
11182 & 2.878 & 10.36 & 3241 & 4.855 & $8.915 \cdot 10^{30}$ & 6 & 1 \\
33905 & 2.572 & 10.09 & 3394 & 4.854 & $1.218 \cdot 10^{31}$ & 4 & 2 \\
34900 & 1.341 & 6.363 & -- & -- & -- & 10 & 1 \\
\dots & \dots & \dots & \dots & \dots & \dots & \dots & \dots\\
\hline
\end{tabular}
\end{table}

\begin{table}
\caption{The final flare catalog. In the case of unsuccessful extraction, we omit the last four columns. The full table is available online.}
\label{table:tess_flares}
\centering
\tiny
\tabcolsep=0.11cm
\begin{tabular}{ccccccc}
\hline \hline
TIC & sector & $t_{\rm peak}$ & $t_{1/2}$ & $A$ & ED & $E_{\rm TESS}$\\
 & & [TBJD] & [min] & & [min] & [erg]\\
\hline
114953216 & 1 & 1342.856475 & 3.649 & 0.02447 & 0.16772 & $3.39 \cdot 10^{31}$ \\
114953216 & 1 & 1345.296690 & 3.246 & 0.06593 & 0.42157 & $8.53 \cdot 10^{31}$ \\
114953216 & 1 & 1345.909173 & 6.421 & 0.09634 & 0.90681 & $1.83 \cdot 10^{32}$ \\
\dots & \dots & \dots & \dots & \dots & \dots & \dots\\
\hline
\end{tabular}
\end{table}

\begin{figure}[ht]
\includegraphics[width=\columnwidth]{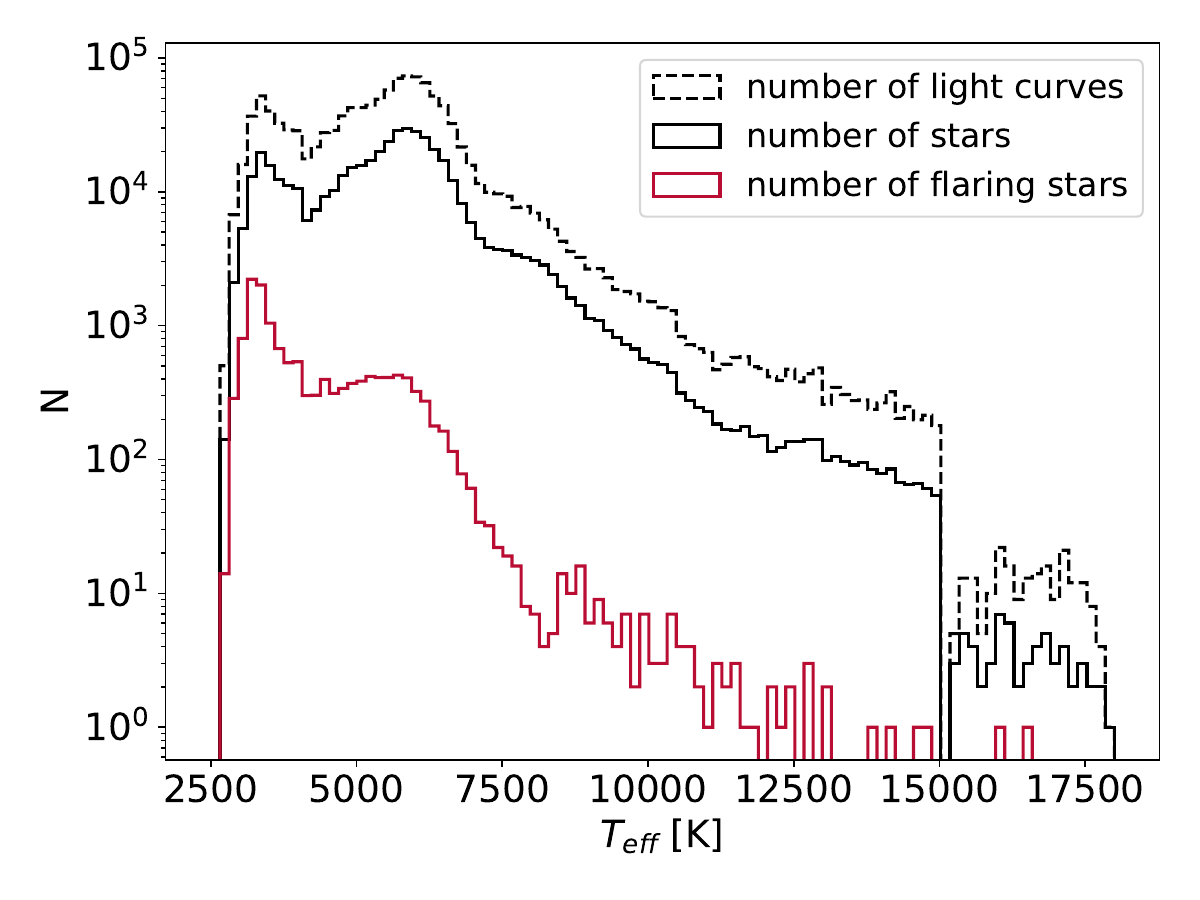}
\caption{Histogram of stars observed with TESS 2-min cadence up to sector 69. The \teff values are from TICv8.2. The distribution is truncated at 18,000\,K.}
\label{fig:numbers_hist}
\end{figure}

Figure~\ref{fig:numbers_hist} shows the \teff distribution of the sample, compared to the \teff distribution of all the stars observed by TESS (9\% of which have no \teff in TICv8.2). Since the number of detectable flares depends on the total observing time, the number of one sector long light curves is also shown. It can be seen that the observed flare rate declines with \teff. However, the target selection for TESS 2-min cadence observations is not random, as stressed by \cite{2020AJ....159...60G}.

\begin{figure}[ht]
\includegraphics[width=\columnwidth]{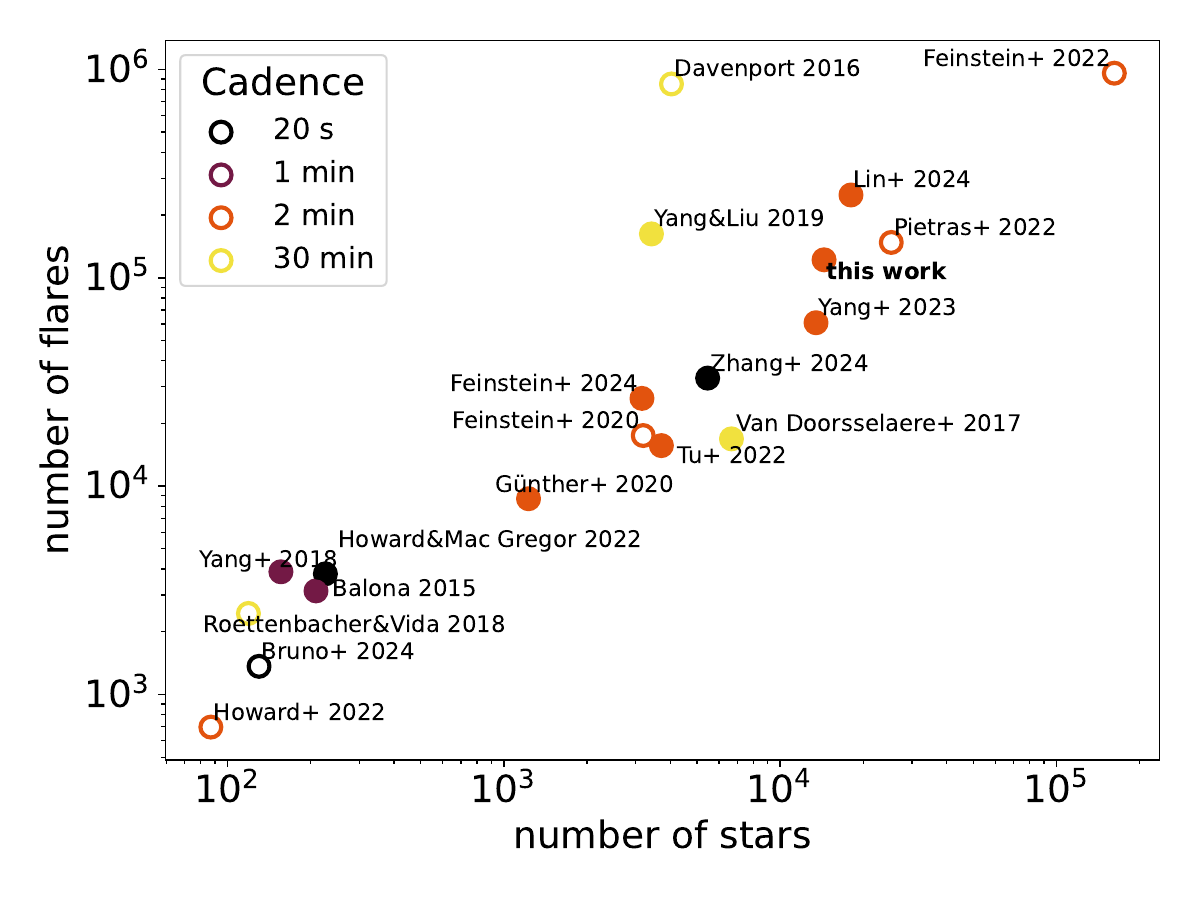}
\caption{Sample size comparison between different stellar flare catalogs created from \kepler{} and TESS data. Color indicates the observing cadence. Filled circles are catalogs that are publicly available. The following catalogs are shown: \cite{2015MNRAS.447.2714B, 2016ApJ...829...23D, 2017ApJS..232...26V, 2018ApJ...868....3R, 2018ApJ...859...87Y, 2019ApJS..241...29Y, 2020AJ....160..219F, 2022ApJ...925L...9F, 2024AJ....168...60F, 2020AJ....159...60G, 2022MNRAS.512L..60H, 2022ApJ...926..204H, 2022ApJ...935..143P, 2022ApJ...935...90T, 2023A&A...669A..15Y, 2024A&A...686A.239B, 2024AJ....168..234L, 2024A&A...689A.103Z}.}
\label{fig:flare_catalog_summary}
\end{figure}

Figure~\ref{fig:flare_catalog_summary} puts the sample size into context, by showing other published flare catalogs. In the lower-left corner we find more focused lists (e.g., TOIs, superflare stars), while in the upper-right corner there are more general catalogs.

\begin{figure}[ht]
\includegraphics[width=\columnwidth]{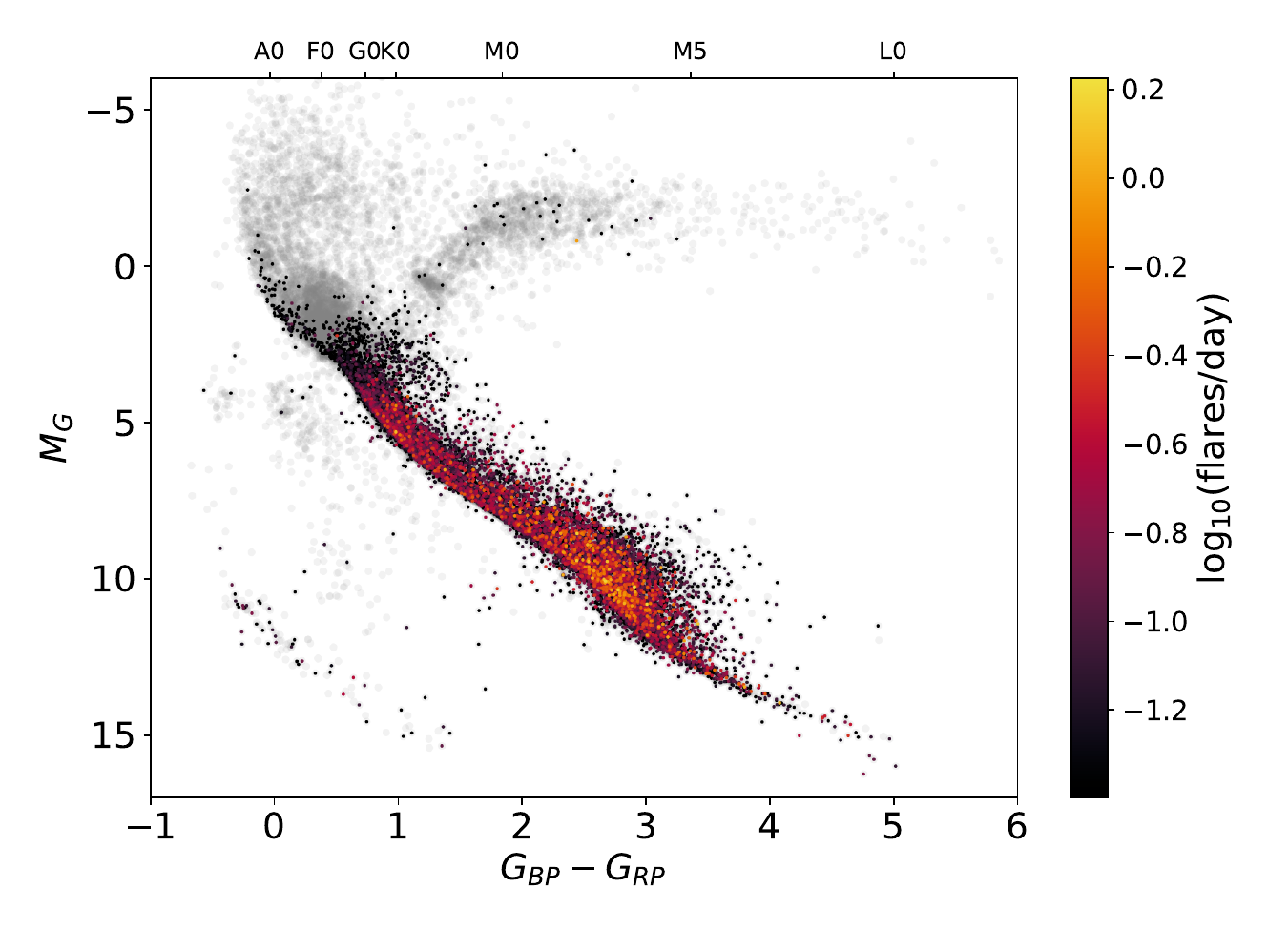}
\caption{The flaring stars on the Gaia color--magnitude diagram, colored with the flare rate. Note that the stars are plotted in order of their flare rates, to show the most active stars on top. Gray points show all the stars prior to manual vetting (Sect.~\ref{sect:manual_vetting}), to make the position of the red giant branch more discernable.}
\label{fig:hrd}
\end{figure}

Figure~\ref{fig:hrd} shows the Gaia color--magnitude diagram of the flaring sample. While there are a few flaring stars on the red giant branch and also a few among white dwarfs, most of them are on the main sequence (MS). The majority of the sample consists of M-dwarfs, and the rate of activity declines for earlier type stars on the MS. The unresolved binary MS is prominent, $\dotem{0.75}$ above the MS. Most stars only have a few flares in the catalog, the median is 3 flares per star, and only 10\% of the stars have more than 20 flares. The three stars with the largest number of detected flares are TIC~150359500, 272232401 and 220433364, with over 500 flares each. Most stars in the sample are nearby: 50\% are closer than 92\,pc, and 90\% are closer than 202\,pc.

\begin{figure*}[t]
\centering
\includegraphics[width=2\columnwidth]{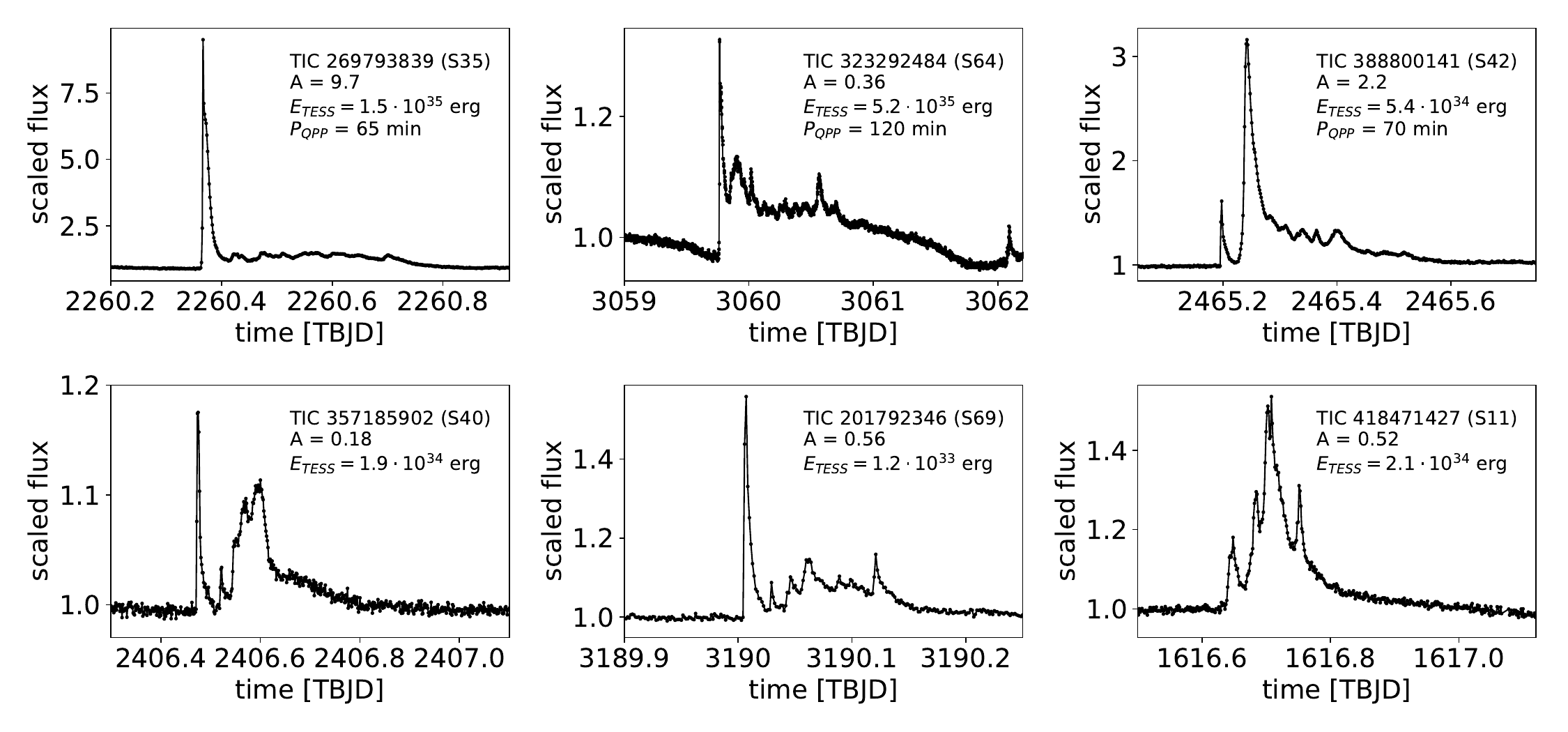}
\caption{Some interesting complex flares identified during manual vetting. The upper panels show flares with possible quasi-periodic modulation.}
\label{fig:hall_of_fame}
\end{figure*}

Figure~\ref{fig:hall_of_fame} shows six interesting flares found during the manual vetting. They all have large amplitudes and show complex behaviour, including quasi-periodic modulation. These events are also relatively long, the flare on TIC~323292484 lasted for more than a day.

\subsection{Binning stars on the color--magnitude diagram}
\label{sect:binning_method}

In the following sections, we explore the possibility of systematic variation of flare properties with stellar astrophysical parameters. For this, we need to place the flares into bins of stellar parameters.

We group stars into bins along the MS, based on the Gaia color--magnitude diagram. Since almost all stars in the sample are on the MS (see Fig.~\ref{fig:hrd}), most of the parameters that are available in bulk are tightly correlated (e.g., \teff, $\log g$, luminosity, radius). Thus, we only show the effect of a single parameter, as many others would lead to similar results, and it is hard to distinguish which is the parameter that directly causes the change.

\begin{figure}[h]
\includegraphics[width=\columnwidth]{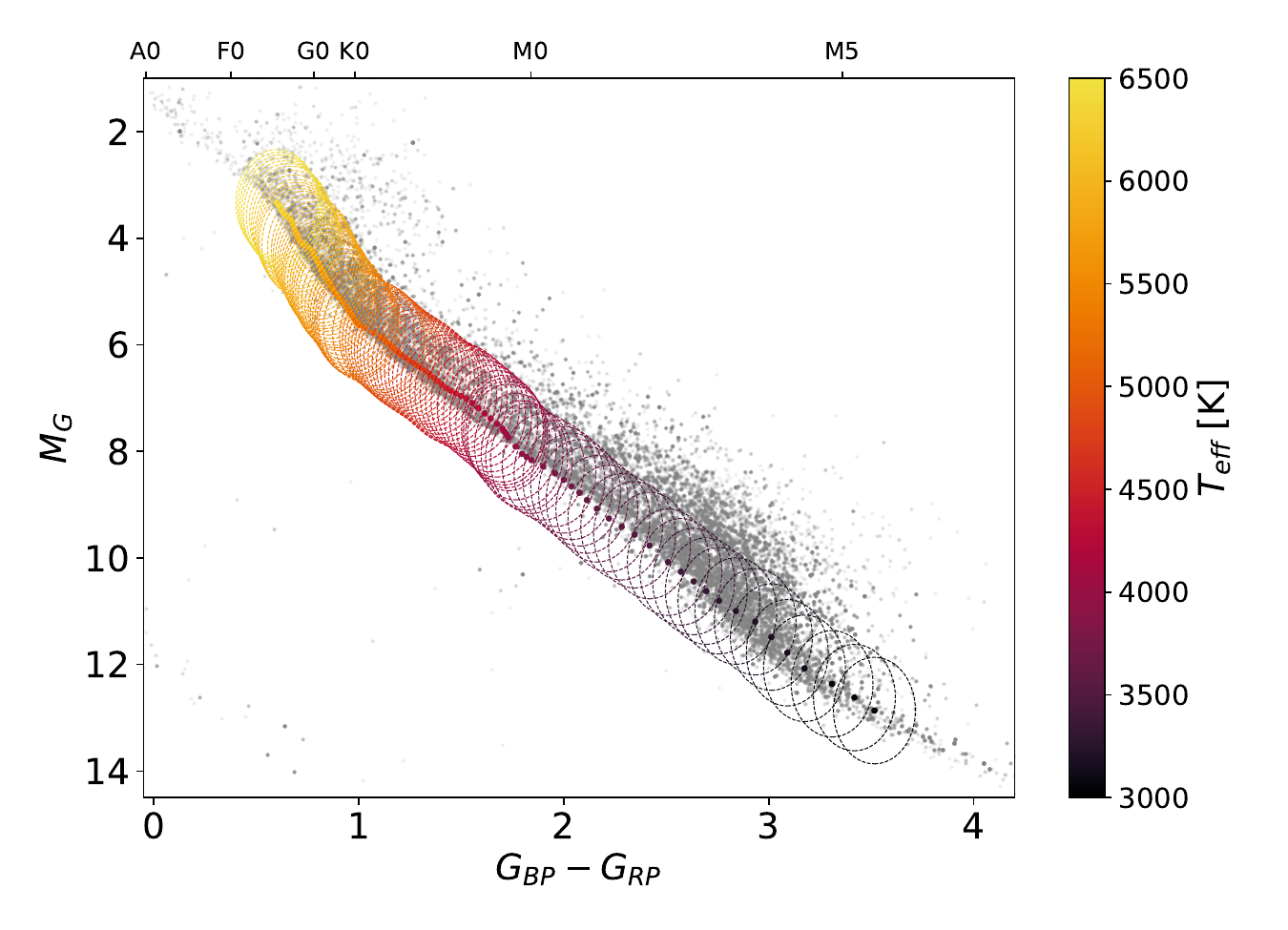}
\caption{Binning on the Gaia color--magnitude diagram for the calculation of the average flare shapes. Around each point, the stars inside an ellipse are counted.}
\label{fig:hrd_sequence}
\end{figure}

To trace the MS, we use the \textit{"Modern Mean Dwarf Stellar Color and Effective Temperature Sequence"}\footnote{Version 2022.04.16 retrieved from \url{https://www.pas.rochester.edu/~emamajek/EEM_dwarf_UBVIJHK_colors_Teff.txt}} from \cite{2013ApJS..208....9P}. We parameterize the sequence with \teff, but the binning is performed in the color--magnitude space, similar to the sample selection in \cite{2021A&A...650A.138S}. We chose this approach over binning by the \teff values from TIC, as those are compiled from different sources, and are thus less homogeneous. First, we interpolate the sequence to 100 uniform \teff values between 3000 and 6500\,K. Then, for each \teff value, we linearly interpolate the corresponding Gaia $(G_{BP}-G_{RP})_i$ color and $M_{G,i}$ absolute magnitude from the sequence, and draw an ellipse on the color--magnitude diagram as follows:
\begin{equation}
    \left( (G_{BP}-G_{RP}) - (G_{BP}-G_{RP})_i \right)^2 + \left( \frac{M_G - M_{G,i}}{5} \right)^2 < 0.2^2
\end{equation}
The stars inside this ellipse make up the sample in the $i$th bin. Figure~\ref{fig:hrd_sequence} illustrates this binning procedure, which we use for the basic flare parameters and the average flare shapes. In the following -- where applicable -- we use the color coding from Fig.~\ref{fig:hrd_sequence}.

\subsection{Basic flare properties}

Before analyzing the non-parametric flare shapes, we examine the following simple parameters estimated for each flare: $A$, \thalf, ED, and energy.

The \thalf distribution of the TESS flares follows a log-normal distribution with location and scale parameters $\mu = 0.83$ and $\sigma = 0.21$ in minutes, giving a median \thalf of 7 minutes. The distribution is truncated at 2 minutes due to our selection criteria described in Sect.~\ref{sect:extraction}.

\begin{figure*}[t]
\centering
\includegraphics[width=1.9\columnwidth]{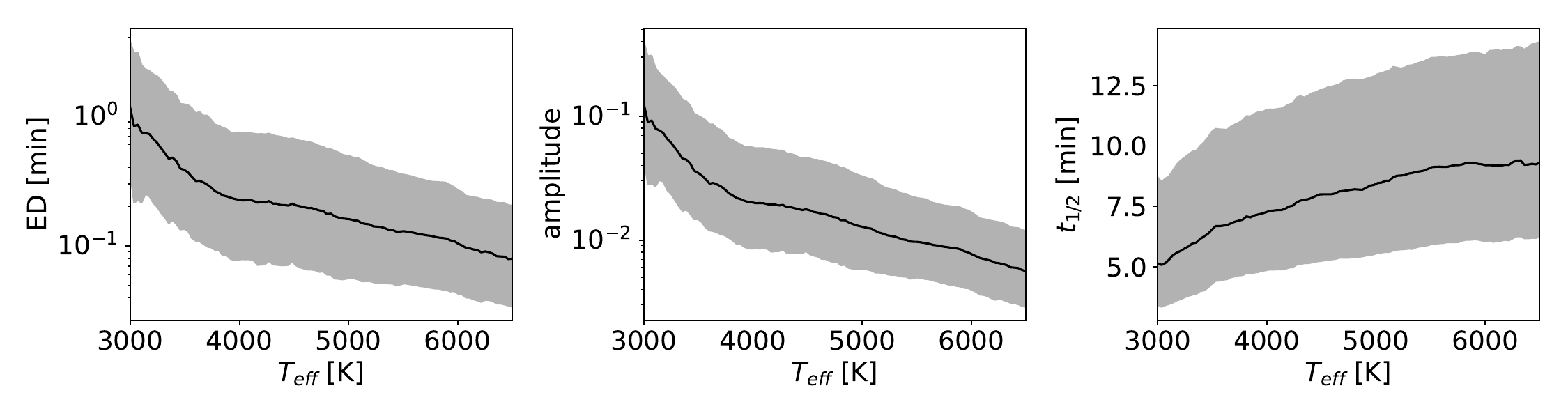}
\caption{The change of basic flare parameters across the MS, with the same binning as on Fig.~\ref{fig:hrd_sequence}. The black line shows the median value and gray shading is shown between the 16th and 84th percentiles.}
\label{fig:t12_ampl_ED_Teff}
\end{figure*}

Figure~\ref{fig:t12_ampl_ED_Teff} shows how the basic flare parameters change across the MS, using the binning described in Sect.~\ref{sect:binning_method}. We argue that the correlation between \teff and ED or amplitude (as seen in e.g., \citealt{2023A&A...669A..15Y} or \citealt{2023MNRAS.523.2193L}) could arise due to the increasing luminosity with \teff on the MS, as the contrast of the flare depends on the quiescent luminosity of the star. Also, flares with smaller amplitude and ED are more numerous on any star, as they follow a power-law distribution \citep[see e.g.,][]{2024LRSP...21....1K}. However, due to the larger photometric errors on the fainter late type stars and the finite time resolution, the distribution of detectable flares will peak at a given ED and amplitude for stars with different \teff. The relationship between \thalf and \teff is also probably attributed to a sampling bias. As the \teff increases, only the more energetic flares can be detected, which will have longer durations \citep[see e.g., ][]{2015EP&S...67...59M, 2017ApJ...851...91N}. However, it is also consistent with the results of \cite{2015MNRAS.447.2714B}, \cite{2020A&A...641A..83K} and \cite{flaring_giants}, who found that flares on giant stars last longer than on the MS, as the flare duration increases with the stellar radius. Also, \cite{2023ApJ...958....9R} showed that the flare decay time scales with the loop length in the corona, which scales with the radius of the star. On the MS, the stellar radius increases with \teff, so the right panel of Fig.~\ref{fig:t12_ampl_ED_Teff} can also be interpreted as \thalf increasing with the stellar radius. However, as there are only a few bonafide giant stars in our sample, it is not possible to break the degeneracy and pinpoint the underlying variable behind the correlation.

Using our homogeneously estimated flare parameters, we can study the relationship between $A$, \thalf and ED. The template provided by \cite{davenport_template} can be analytically integrated to calculate the ED as follows:
\begin{equation}\label{eq:Davenport_ED}
	\mathrm{ED}(A, t_{1/2}) = 1.827 \cdot A \cdot t_{1/2}
\end{equation}
As a more general form, we fit the following power law to the data in bins of stellar parameters:
\begin{equation}\label{eq:ED_from_A_t12}
	\mathrm{ED}(A, t_{1/2}) = \alpha \cdot A^{\beta} \cdot t_{1/2}^{\gamma}
\end{equation}
For the case of the analytic template of \cite{davenport_template}, $\alpha=1.827$, $\beta=1$, and $\gamma=1$.

\begin{figure*}[t]
\centering
\includegraphics[width=1.9\columnwidth]{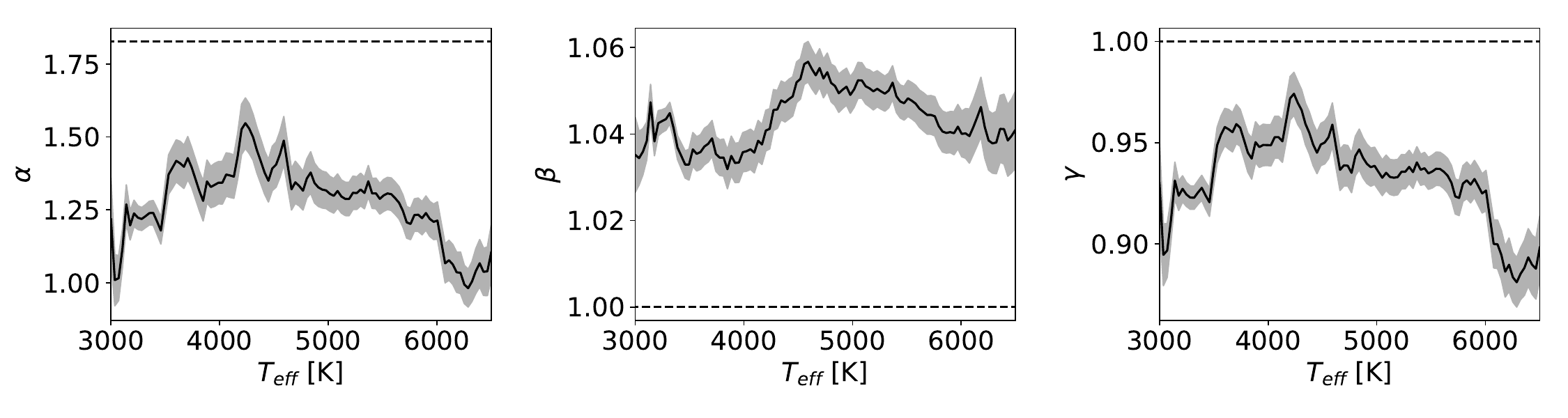}
\caption{Fitted parameters for the power law in the form ED$(A, t_{1/2}) = \alpha \cdot A^{\beta} \cdot t_{1/2}^{\gamma}$, with the same binning as on Fig.~\ref{fig:hrd_sequence}. The gray area denotes the uncertainty of the fit. The dashed lines show the parameters for the analytic template of \cite{davenport_template}.}
\label{fig:basic_flare_parameter_fits}
\end{figure*}

Figure~\ref{fig:basic_flare_parameter_fits} shows how the fitted parameters change along the MS, using the binning described in Sect.~\ref{sect:binning_method}. It hints at systematic differences in the flare shape, and a more complex ED$(A, t_{1/2})$ relationship than Eq.~\ref{eq:Davenport_ED}.

\subsection{Flare shape space}

\begin{figure}[h]
\includegraphics[width=\columnwidth]{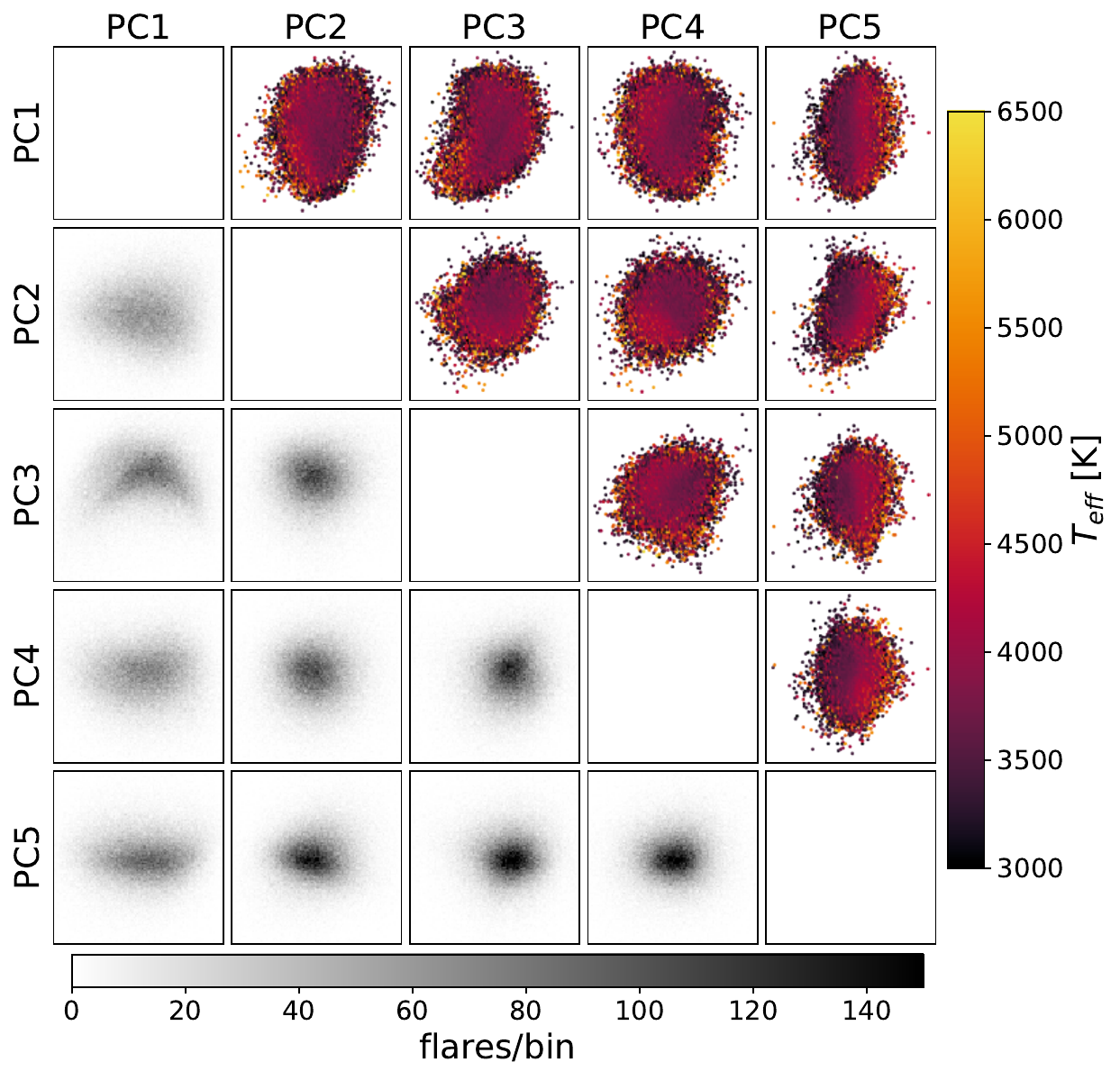}
\caption{Distribution of flares in the principal component space. Each panel shows a two-dimensional histogram. Below the diagonal, the shading indicates the number density of flares. Above the diagonal, the color code indicates the average \teff from TICv8.2 in each bin. The \teff dependence is most apparent in PC5.}
\label{fig:pc_space}
\end{figure}

Each flare in the sample was extracted between $-3$ and 10\thalf, and contained tens to hundreds of data points in this region, with a median of 44 points. They were then linearly interpolated to 200 points, making the scaled flare shape dataset 200 dimensional, which is too high for visualization. We used WPCA to transform the data into a lower-dimensional space. Figure~\ref{fig:pc_space} shows a 5-dimensional representation of the dataset, with 2-dimensional projections. The distribution in this principal component (PC) space appears to be smooth, and the distributions of higher PCs appear to be more symmetrical. {This symmetry suggests that the higher PCs only describe random noise, and are not physically interesting.}

\begin{figure*}[h]
\includegraphics[width=2\columnwidth]{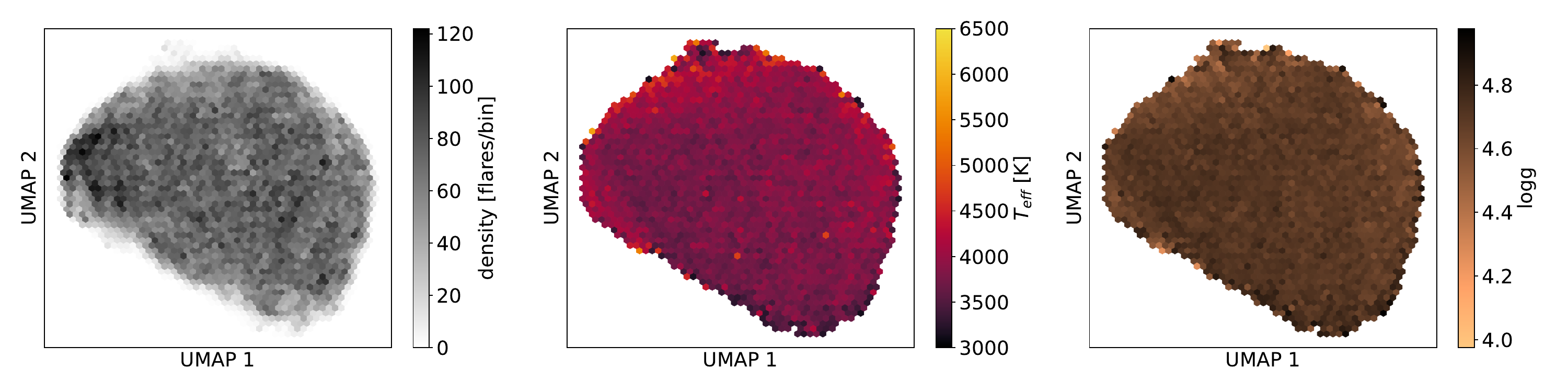}
\caption{Two-dimensional UMAP projection of the scaled flare shapes. Each panel shows a two-dimensional histogram, color-coded by the density of points, \teff and $\log g$ from TICv8.2.}
\label{fig:umap}
\end{figure*}

Figure~\ref{fig:umap} shows the 2-dimensional representation of the dataset with UMAP. The distribution is again really smooth, with no clear sign of clustering.

We briefly mention that these 2-dimensional histograms are not the only way to visualize the flare shape space. \cite{2022MNRAS.515.3883G} visualized the parameter space of 2500 pulsars as a graph, using the minimum spanning tree. This way, similar objects can be grouped close to each other, and the structure of the graph can be analyzed.

\subsubsection{Possible multimodality}

One way to look for distinct flare shapes is to find clusters in a lower dimensional representation, e.g., the PC space. Each of the marginal distributions in Fig.~\ref{fig:pc_space} appears to be unimodal, and to test it, we can use different clustering algorithms.

First, we use Hierarchical Density-Based Spatial Clustering of Applications with Noise (HDBSCAN, \citealt{Campello2013DensityBasedCB, McInnes2017}) on the first 20 PCs. \mbox{HDBSCAN} is a sophisticated clustering algorithm that aims to identify clumps in the data above a less dense background. 
It is able to separate the optimal number of clusters automatically, without the need to specify the number. 
It does not assign every data point to a cluster, as in the case of partitioning algorithms like $k$-Means. 
It is frequently used for finding open clusters and associations surrounded by field stars in Gaia data \citep[see e.g.,][]{2023A&A...673A.114H}.
By requiring at least 10 points in each cluster, \mbox{HDBSCAN} only found a single cluster above the background, at the core of the distribution, indicating no signs of multimodality. The result is the same when using only 5 PCs.

Then, we applied Gaussian mixture models \citep{2014sdmm.book.....I} to the first 5 PCs. This method is a density estimation technique that fits the data with a given number of N-dimensional Gaussians. It can also be used for soft clustering, assigning each data point to clusters probabilistically. The number of Gaussian components can be selected using BIC. By varying the number of Gaussians between 1 and 30, the preferred values ranged from 15 to 20, with slight differences in BIC. However, these are not distinct clusters, but rather overlapping ones. We calculated the silhouette score \citep{ROUSSEEUW198753} for different number of Gaussians, and found values below 0.1, which hints to the absence of distinct clusters. Silhouette scores close to one indicate strong, non-overlapping clumps, while values close to zero indicate totally overlapping clusters.

\begin{figure}[h]
\includegraphics[width=\columnwidth]{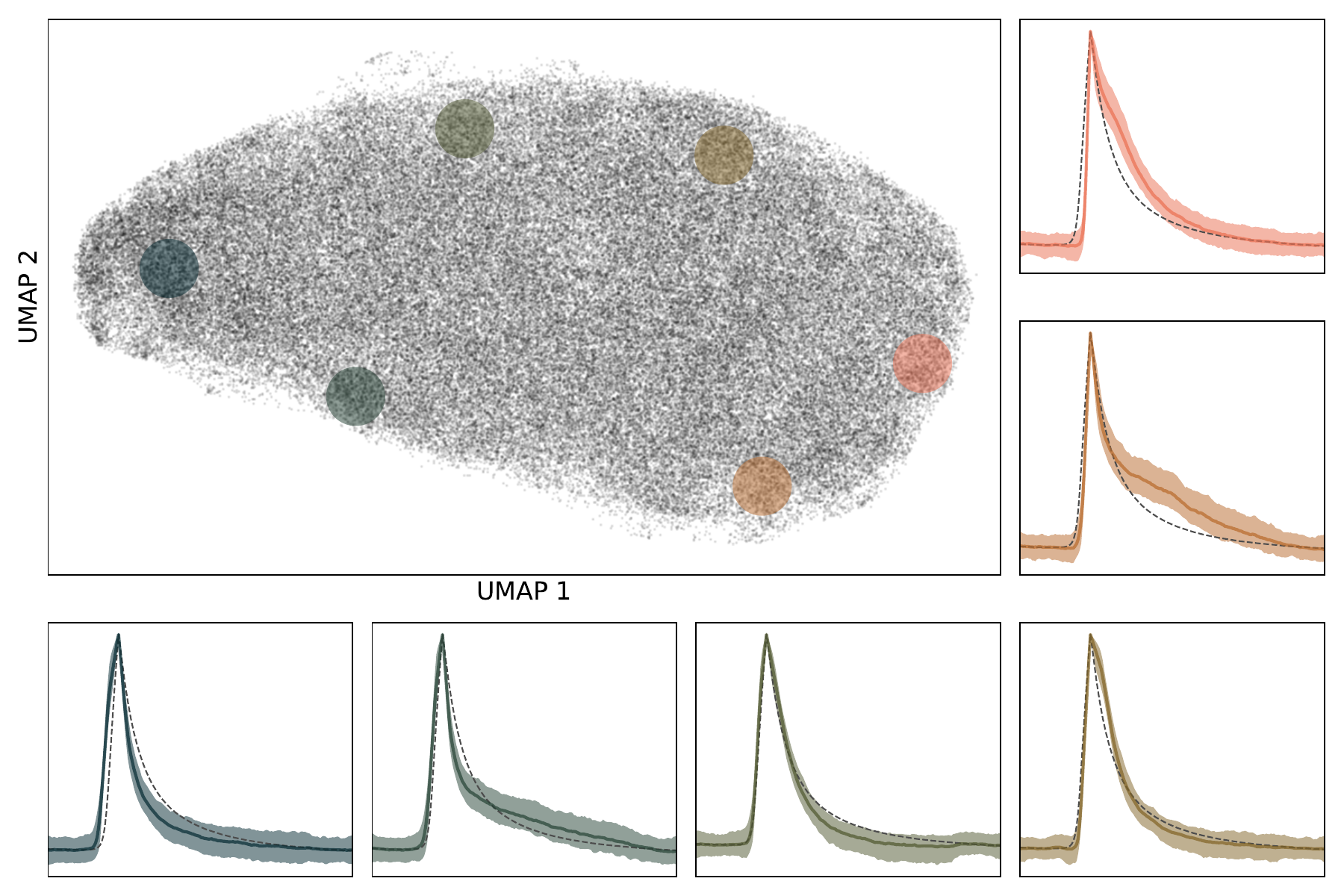}
\caption{Average flare shapes from different positions in the UMAP space. Different colors show the median profiles and the range between the 16th and 84th percentiles inside the given circles in the UMAP space. Each circle includes approximately 1000 flares. Dashed lines in the subplots show the whole sample median for comparison.}
\label{fig:umap_example_shapes}
\end{figure}

While these methods do not prove that flare shape clusters do not exist, they suggest a more gradual, continuous change in the scaled flare profiles. Figure~\ref{fig:umap_example_shapes} illustrates that the shapes indeed change in the dimensionality reduced space, by showing the median profiles from six different regions in the UMAP space.

\subsection{Changing flare shapes with astrophysical parameters}

Figure~\ref{fig:pc_space} already hinted to a systematic variation of flare shapes with \teff, with a clearly visible color gradient mostly in PC5. Before attempting to extract astrophysical information from \textit{individual} flares, we consider the changes in the \textit{average} shape of a large number of flares from similar stars. For this, we bin the stars by their position on the Gaia MS, as described in Sect~\ref{sect:binning_method}.

\begin{figure}[t]
\includegraphics[width=\columnwidth]{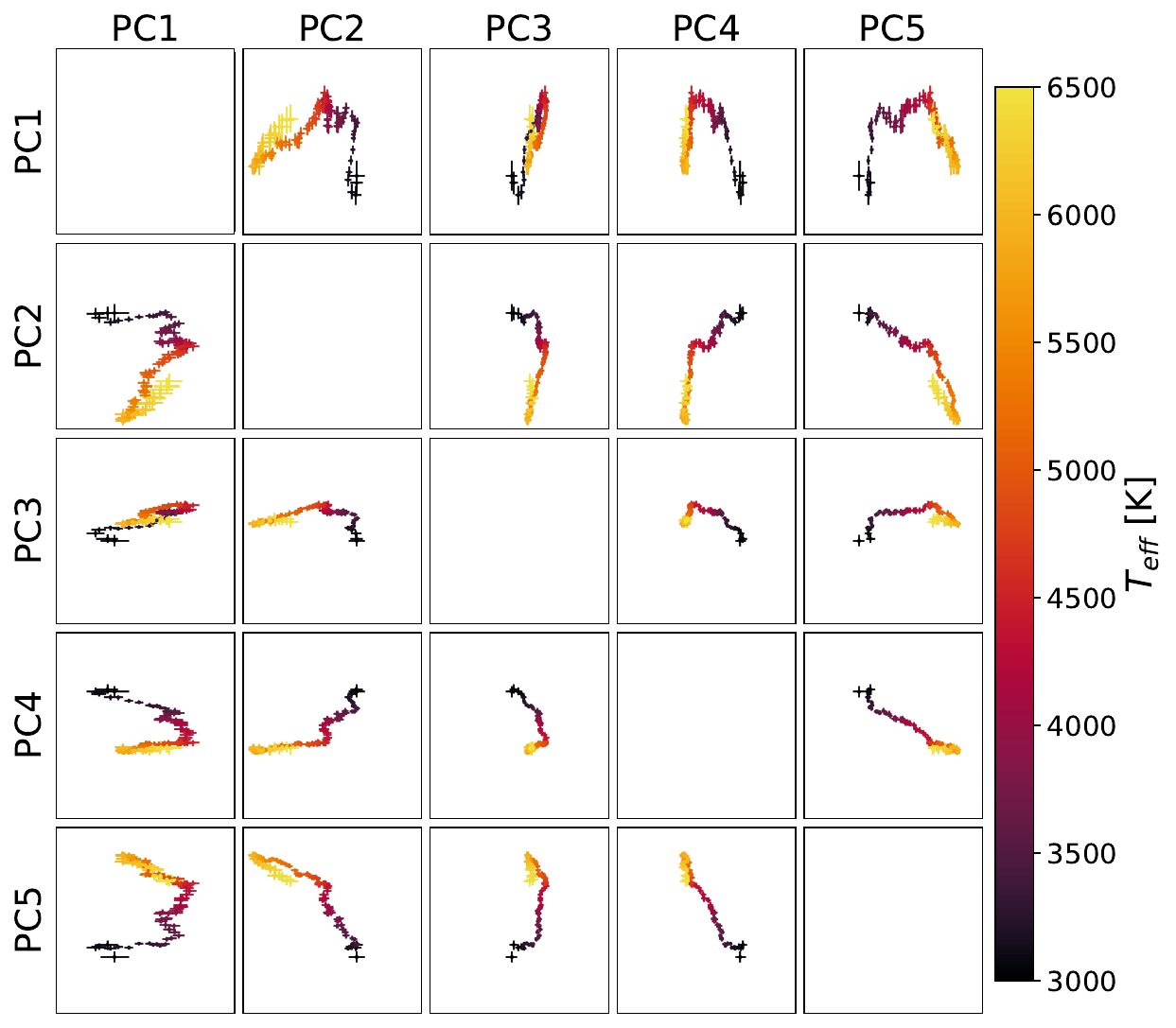}
\caption{Changes in the flare shapes in the principal component space, along the MS. The points denote the median value of the PC coefficients for the stars in the given bins from Fig.~\ref{fig:hrd_sequence}, colored with \teff. The error bars show the standard error of the median, $\sqrt{\frac{\pi}{2N}}$ times the standard deviation (assuming roughly normal distributions). In each subpanel, only the inner 10\% of the range from Fig.~\ref{fig:pc_space} is shown.}
\label{fig:pc_space_sequence}
\end{figure}

Figure~\ref{fig:pc_space_sequence} shows the position of the median flare shape in the PC space for each bin, color-coded with \teff. There is a systematic "wandering" of the points in the PC space, following the color gradients in Fig.~\ref{fig:pc_space}. The simplest trajectories with the least turns lie along PC5, indicating that it is directly proportional to \teff. The Pearson correlation coefficient between \teff and PC5 is 0.15 with $p<10^{-200}$, which is the strongest correlation among the PCs.

\begin{figure*}[h]
\centering
\includegraphics[width=2\columnwidth]{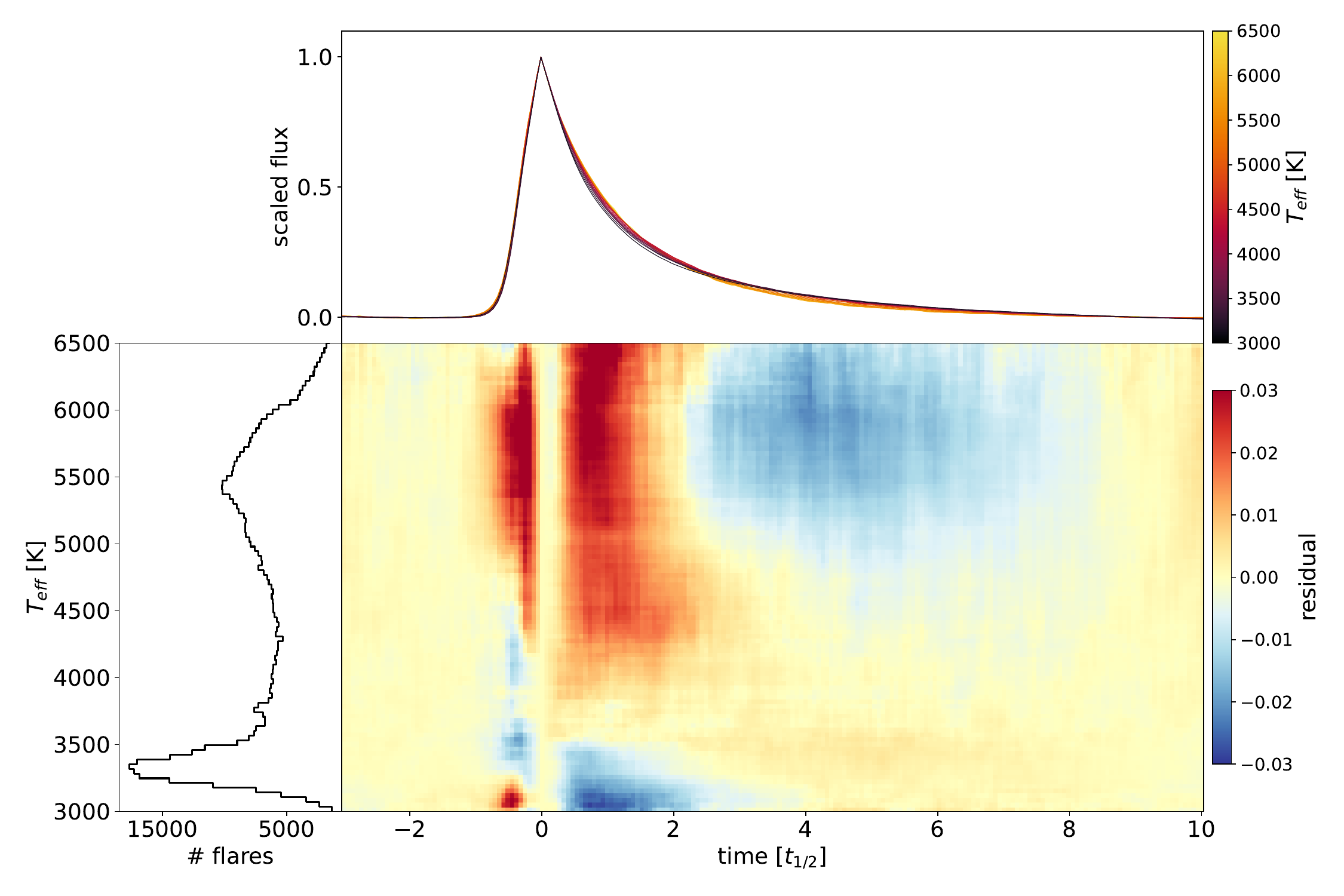}
\caption{Flare shapes along the MS, using the binning from Fig.~\ref{fig:hrd_sequence}. \textit{Top:} Median shapes colored with \teff. \textit{Middle:} Residual flare shapes, made by removing the whole sample median from the median shape in each bin. \textit{Left:} Number of flares in each bin along the MS, parameterized by \teff.}
\label{fig:shapes_sequence}
\end{figure*}

Figure~\ref{fig:shapes_sequence} shows how the median flare shape changes along the MS. The upper panel shows the scaled flare shapes, and the lower panel shows the residual, after removing the median flare shape of the whole sample. There are striking variations in the shape, albeit with relatively small, few percent amplitude. The main feature is that the flares of hotter stars are "fatter", and wider for a few \thalf, but they decay more quickly after $\sim2$\thalf. The zero residuals around $-3$\thalf and $10$\thalf are artifacts of the flare extraction, caused by the final linear detrending, which is performed on those regions.

To explore whether we could have revealed any other type of flare shape variability, we present mock retrieval tests in Appendix~\ref{app:mock_flares}. We inject three different kinds of trends into real light curves and find that the only kind of variability that we can safely recover is similar to what we have found in Fig.~\ref{fig:shapes_sequence}.

\subsection{Astrophysical parameter prediction from individual flares}
\label{sect:predictions}

To find nontrivial relationships between astrophysical parameters and flare shape, we can also use flexible machine-learning models. If we can find an algorithm that predicts a certain astrophysical parameter of the star only from the shape of its flares, we can prove that such a dependence exists.

To parameterize the flare shapes, we use 20 components from WPCA and also add the \thalf, amplitude, and ED. We use this 20+3 dimensional dataset as input and \teff from TICv8.2 as output labels. We remove stars with $T_{\mathrm{eff}}>8000$\,K. We use the base-10 logarithm of \thalf, amplitude and ED, as they change multiple orders of magnitude. Due to missing \teff values and some negative EDs, we are left with a sample size of 108,687. Before training, we subtract the median from each feature, and scale by the interquartile range to normalize the data. We train different regression models and compare the results to a dummy regressor that outputs the mean label for each input. To assess the performance of the models, we use $k$-fold cross-validation with root mean squared error as a metric. We split the dataset into $k=5$ partitions (folds), and use one of them as the test set, and the other four as the training set. The final metric is calculated as the average of the folds.

We try the following regression models, using their \texttt{scikit-learn} \citep{scikit-learn} implementations: multivariate linear regression, random forest \citep{10.1023/A:1010933404324} and gradient boosting \citep{10.1214/aos/1013203451}. Random forest is an ensemble method that averages the results of multiple decision trees, where each tree has the same number of branches (depth), splitting the input data by a single feature at each branch. Gradient boosting also works with a given number of trees, but the results of the trees are not averaged, instead, they are built atop of each other, boosting the performance of previous trees. We use the following fine-tuned parameters for the models: 20 trees with depth=10 for random forest, and 20 trees with depth=10 for gradient boosting.

Table~\ref{table:pred_results} summarizes the results. For the prediction of \teff, the best model outperforms the dummy regressor by only 5--30\%. The prediction accuracy improves when the scaling parameters \thalf, $A$ and ED are included. The best-performing models are the random forest and gradient boosting with similar scores.

Based on these results, we conclude that the shapes of \textit{individual} flares do not carry enough information to determine the astrophysical parameters of their stars, although there is a marginal improvement over the dummy regressor even in the PCA-only case. Adding the flare amplitude as input improves the accuracy, due to the different flare contrast on stars with different luminosities. By shuffling the data points of single input features, the permutation feature importance score can be calculated, i.e., how much the regressor relies on a given input feature. Based on this, the most important feature is $A$, followed by \thalf, ED, PC2 and PC5.

We also tried to average the flares for each star, to see whether it improved the regression performance. We could achieve only a few percent improvement in root mean squared error compared to the best model using individual flares, and only when the number of flares per star was included as an input parameter. For reliable prediction of \teff, we estimate that one would need to average at least a few hundred flares for each star (as in the case of Fig.~\ref{fig:shapes_sequence}), which is only available for a handful of objects.

\begin{table}[ht]
\caption{Root mean squared error for \teff from the $k$-fold cross-validation, using only the PCA components, only \thalf, $A$ and ED, and all of them as input features.}
\label{table:pred_results}
\centering
\small
\begin{tabular}{c|ccc}
\hline
 & \multicolumn{3}{c}{$\Delta$\teff [K]} \\
 Model & PCA & \thalf, $A$, ED & both \\
\hline
Dummy regression & $857 \pm 37$ & $857 \pm 37$ & $857 \pm 37$ \\
Linear regression & $837 \pm 35$ & $683 \pm 20$ & $675 \pm 20$ \\
Random forest & $826 \pm 35$ & $\mathbf{643 \pm 21}$ & $612 \pm 19$ \\
Gradient boosting & $\mathbf{825 \pm 35}$ & $648 \pm 23$ & $\mathbf{608 \pm 21}$ \\
\hline
\end{tabular}

\end{table}

\section{Application of the results}
\subsection{New flare templates}

Using a large number of high-quality, manually selected 1-min cadence \kepler{} flares from the M4 dwarf GJ~1243, \cite{davenport_template} created a flare shape template that has been used in many cases since its release. Here, we aim to study how this flare template would change for different types of stars.

We adopt the same parameterization as \cite{davenport_template}, using a 4th-order polynomial for the rise phase, and the sum of two exponentials for the decay phase, as follows:
\begin{gather}
    F_\mathrm{rise}(t) = 1 + a_1 \cdot t + a_2 \cdot t^2 + a_3 \cdot t^3 + a_4 \cdot t^4 \label{eq:template_rise} \\
    F_\mathrm{decay}(t) = b_1 \cdot e^{-c_1 t} + b_2 \cdot e^{-c_2 t} \label{eq:template_decay}
\end{gather}
We also force both fits to go through the peak ($t=0$, $F=1$), and the rise phase to end at $-$\thalf ($t=-t_{1/2}$, $F=0$). For $t<-$\thalf, the template is zero by definition. We use the scaled and interpolated flares for the fit, grouped together along the MS as described in Sect.~\ref{sect:binning_method}. We use all the flaring points aggregated for the fit, not just an average curve. We make no distinction between simple and complex flares.

\begin{figure}[h]
\includegraphics[width=\columnwidth]{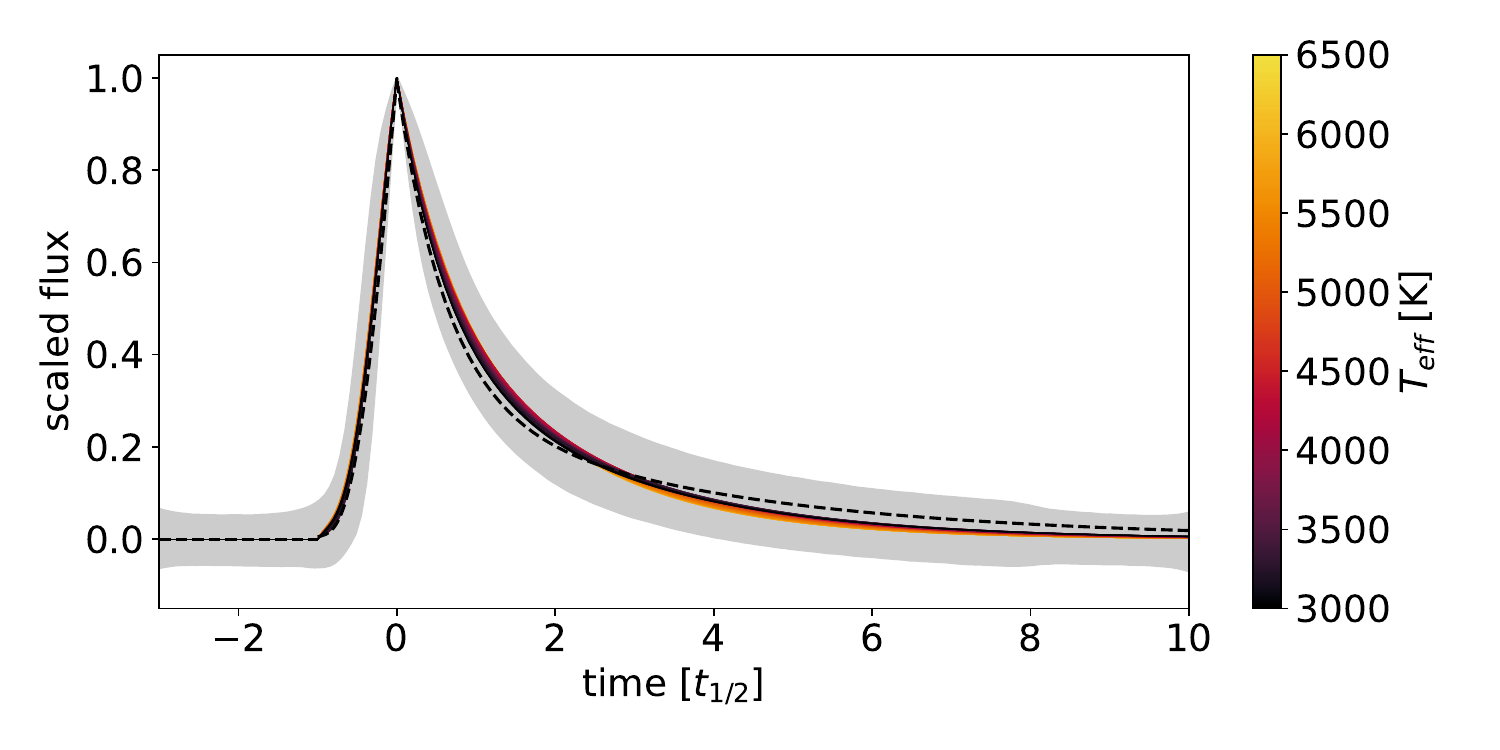}
\caption{Flare templates fitted along the MS, using Eq.~\ref{eq:template_rise}--\ref{eq:template_decay}, color coded with \teff. The shaded region shows the 16th and 84th percentiles of the dataset. The black dashed line shows the template of \cite{davenport_template}.}
\label{fig:template_result}
\end{figure}

\begin{figure*}[h]
\includegraphics[width=2\columnwidth]{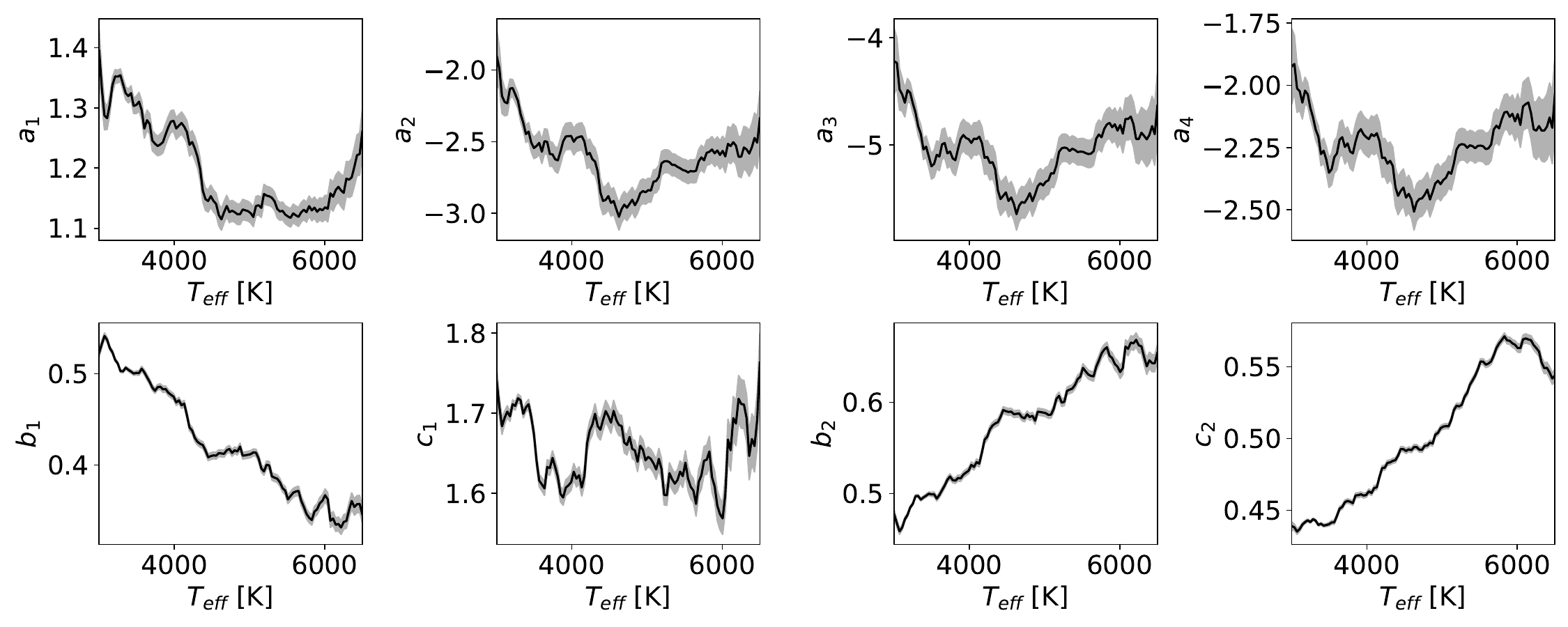}
\caption{Parameters of the flare template fitted along the MS in the following form: $F_\mathrm{rise}(t) = 1 + a_1 \cdot t + a_2 \cdot t^2 + a_3 \cdot t^3 + a_4 \cdot t^4$ and $F_\mathrm{decay}(t) = b_1 \cdot e^{-c_1 t} + b_2 \cdot e^{-c_2 t}$. The shaded regions show the formal uncertainty of the fit.}
\label{fig:template_params}
\end{figure*}

Figure~\ref{fig:template_result} shows these templates plotted for different \teff, along with the original template from \cite{davenport_template}. Since the dataset used to create that template is different in both cadence and passband to our 2-min cadence TESS dataset, any comparison should be made with caution. The templates vary only a few percent for different types of stars, they can broadly recover the trends visible on the residual image in Fig.~\ref{fig:shapes_sequence}. Figure~\ref{fig:template_params} shows how the parameters of the flare template change along the MS. The $c$ parameters have the most physical relevance, as they are the exponents of the exponentials. $c_1$ seems to change erratically, it can be considered constant, while $c_2$, the exponent of the late decay phase changes almost monotonically with \teff. A similar effect was seen in the residual map of Fig.~\ref{fig:shapes_sequence}. The fit is also repeated for broad \teff ranges, and the parameters are reported in Tab.~\ref{table:template_params}.

\begin{table*}[h]
\caption{Parameters of the flare template from Eq.~\ref{eq:template_rise} and \ref{eq:template_decay}, fitted for different \teff ranges. The last row shows the most widely used template made for an M4 dwarf.}
\label{table:template_params}
\centering
\tiny
\tabcolsep=0.11cm
\begin{tabular}{ccccccccc}
\hline \hline
\teff [K] & $a_1$ & $a_2$ & $a_3$ & $a_4$ & $b_1$ & $c_1$ & $b_2$ & $c_2$\\
\hline
3000--3500 & $1.329 \pm 0.006$ & $-2.318 \pm 0.030$ & $-4.838 \pm 0.047$ & $-2.191 \pm 0.024$ &  $0.511 \pm 0.001$ & $1.675 \pm 0.003$ & $0.489 \pm 0.001$ & $0.441 \pm 0.000$ \\
3500--4000 & $1.264 \pm 0.010$ & $-2.589 \pm 0.051$ & $-5.163 \pm 0.082$ & $-2.310 \pm 0.041$ &  $0.484 \pm 0.002$ & $1.624 \pm 0.005$ & $0.516 \pm 0.002$ & $0.453 \pm 0.001$ \\
4000--4500 & $1.248 \pm 0.017$ & $-2.532 \pm 0.087$ & $-5.023 \pm 0.140$ & $-2.242 \pm 0.070$ &  $0.423 \pm 0.003$ & $1.720 \pm 0.011$ & $0.577 \pm 0.003$ & $0.482 \pm 0.002$ \\
4500--5000 & $1.129 \pm 0.018$ & $-2.949 \pm 0.093$ & $-5.542 \pm 0.149$ & $-2.465 \pm 0.074$ &  $0.406 \pm 0.004$ & $1.697 \pm 0.013$ & $0.594 \pm 0.004$ & $0.498 \pm 0.002$ \\
5000--5500 & $1.168 \pm 0.017$ & $-2.595 \pm 0.087$ & $-5.027 \pm 0.140$ & $-2.265 \pm 0.069$ &  $0.362 \pm 0.004$ & $1.726 \pm 0.016$ & $0.638 \pm 0.004$ & $0.539 \pm 0.002$ \\
5500--6000 & $1.132 \pm 0.020$ & $-2.556 \pm 0.101$ & $-4.777 \pm 0.162$ & $-2.089 \pm 0.081$ &  $0.343 \pm 0.006$ & $1.662 \pm 0.021$ & $0.657 \pm 0.006$ & $0.567 \pm 0.003$ \\
6000--6500 & $1.196 \pm 0.035$ & $-2.549 \pm 0.179$ & $-4.922 \pm 0.287$ & $-2.176 \pm 0.142$ &  $0.344 \pm 0.011$ & $1.606 \pm 0.036$ & $0.656 \pm 0.011$ & $0.562 \pm 0.005$ \\
\cite{davenport_template} & $1.941 \pm 0.008$ & $-0.175 \pm 0.032$ & $-2.246 \pm 0.039$ & $-1.125 \pm 0.016$ & $0.6890 \pm 0.0008$ & $1.600 \pm 0.003$ & $0.3030 \pm 0.0009$ & $0.2783 \pm 0.0007$ \\
\hline
\end{tabular}
\end{table*}

\subsection{Sampling flare shapes}

\begin{figure}[h]
\includegraphics[width=\columnwidth]{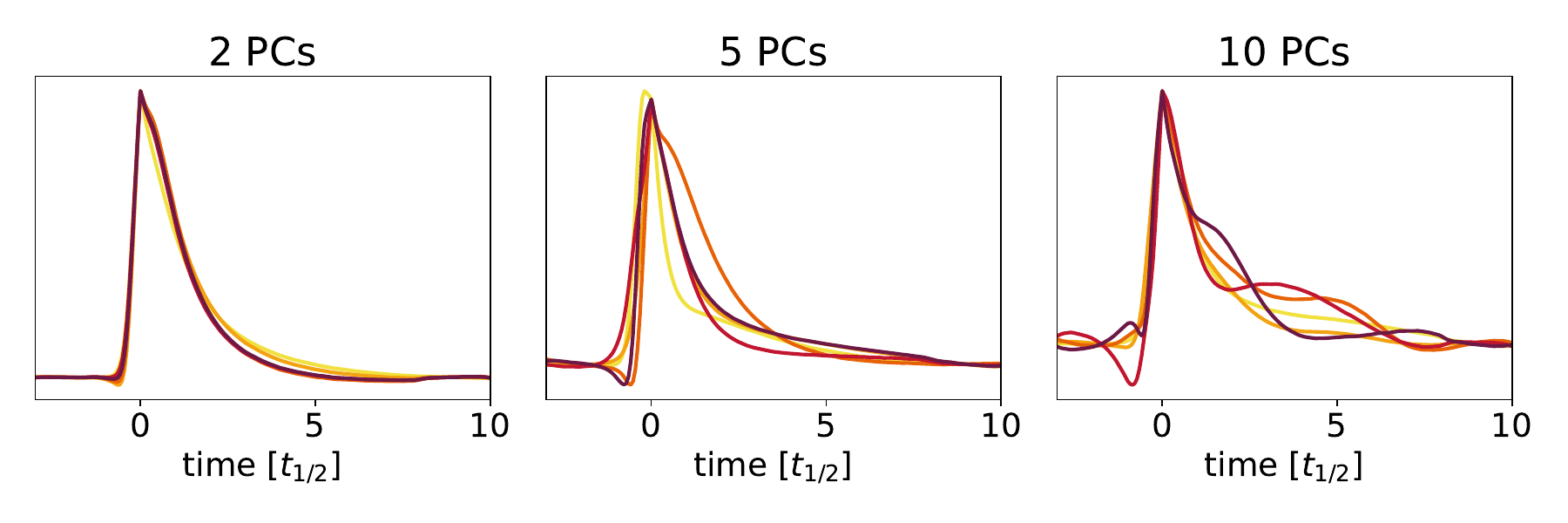}
\caption{Flare shapes randomly sampled from a kernel density estimator trained on the given number of principal components.}
\label{fig:sampling_example}
\end{figure}

When simulating flaring stars' light curves, it is necessary to use realistic flare shapes. Such a situation would arise during the training of data-driven flare detection algorithms, injection-recovery tests, simulating realistic light curves when stellar flares are a source of astrophysical noise, and so on. Apart from using analytical templates, a more sophisticated approach would be to sample from some low-dimensional representation. Such a representation is the PC space of our flare catalog. One could directly sample real flares from the catalog itself, but another solution would be to use a density estimation on the PC space, sample from that, and transform them to flare shapes using the PCA basis. Normalizing flows and variational autoencoders are specifically designed for this task (see, e.g., \citealt{2024MNRAS.533..143L, 2024PhRvD.110l3007S}), however, as the topology of the PC space in this case is simple enough, it is reasonable to use a simpler model. This can either be a Gaussian mixture model, or a kernel density estimator (KDE, \citealt{2017arXiv170403924C}). Once such an estimator is trained, it is fast to sample from it. One important consideration is that we want to generate smooth, noiseless flares, thus we cannot use too many PCs, as the higher PCs mostly describe noise. Figure~\ref{fig:sampling_example} shows random flare shapes drawn from a Gaussian KDE fitted to the 2, 5 and 10-dimensional PC space. Using 10 PCs results in more "wiggly" flares, as the later PCs mostly describe the noise in the data.

It is also possible to sample from a joint PC--physical parameter distribution, including \thalf and amplitude. Restricting the input data to flares from stars with given characteristics (e.g., spectral type, brightness), it is possible to simulate even more realistic flares for special use cases.

To facilitate the use of our flare catalog for sampling, we provide a short Jupyter notebook on Zenodo that demonstrates how synthetic light curves can be generated with flares drawn from a KDE.

\subsection{Locating similar flares}

Another application of the PCA representation is that we can use it to locate real flares that are similar to a given (real or artificial) input shape. The task can be reduced to a nearest-neighbor search in the PC space, after transforming the input shape into this space. A similar technique was presented by \cite{2023PASP..135h4101S}, using autoencoders to find galaxies with similar morphologies.

\begin{figure}[h]
\includegraphics[width=\columnwidth]{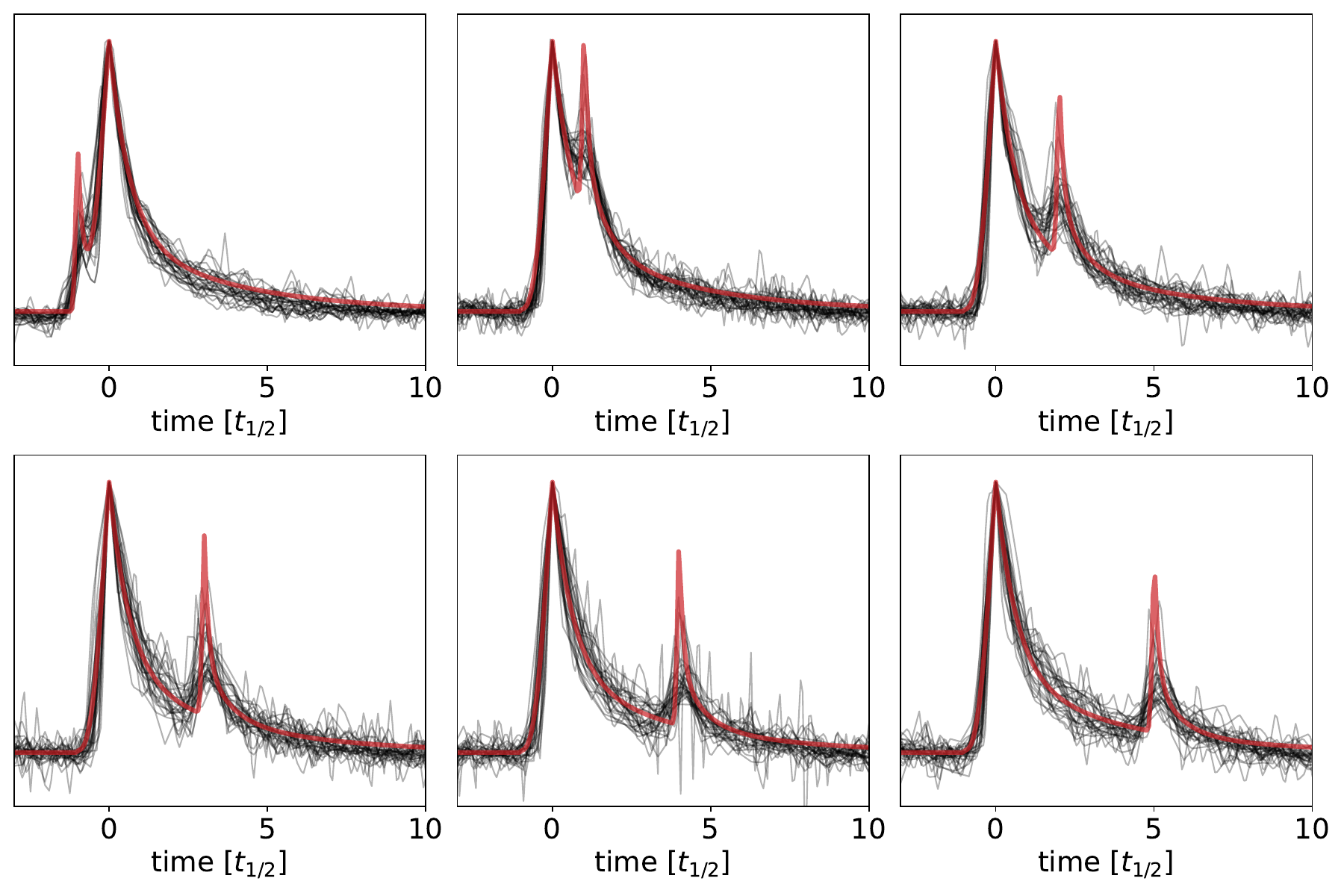}
\caption{Nearest neighbour search in the flare shape space. The red curve on each panel is the injected flare shape, and the 30 closest flares are shown in black.}
\label{fig:PCA_injection}
\end{figure}

Figure~\ref{fig:PCA_injection} shows an example. Double-peaked flares were generated by adding a secondary peak to a template at different positions. These are the input shapes, and we would like to find real flares that look similar to them. For fast nearest neighbour search, a KDTree \citep{1999cs........1013M} was built with Euclidean metric from the first 20 PCs, after normalizing the dataset to have zero mean and unit variance in each dimension. Using the PCA basis, the input shapes were transformed to this space (applying the same scaling). Then, the KDTree was queried with these inputs to locate the 30 closest real flares from the catalog. The red lines in Fig.~\ref{fig:PCA_injection} show the input shapes, and the black lines show the retrieved flares, which are indeed similar to the input.

One important consideration with this method is the number of PCs to keep. On one hand, the search algorithm is affected by the curse of dimensionality. This makes the query less efficient in high dimensional space, where the query reduces to a simple linear search over the whole flare sample \citep{DBLP:reference/cg/2004}. The Euclidean metric is also less relevant in higher dimensions \citep{Aggarwal2001OnTS}. On the other hand, using too few PCs gives poor results, as they cannot describe the input shape adequately. The optimal number of PCs is different for each input shape, depending on which PCs describe the variation the best. After experimenting with different inputs, a number between 10 and 20 seemed adequate.

We provide a short Jupyter notebook on the Zenodo page of the paper that allows the reader to query the catalog for arbitrary input shapes.

\section{Solar flare shapes}

To gain insights into the stellar case, we make a simple attempt to study solar flare shapes. To obtain a dataset comparable to the stellar case, we choose an instrument that provides disk-integrated light curves. As solar white light flares are rare, we use an instrument operating in the ultraviolet regime. To have a hypothesis to test, we compare light curve shapes of flares with and without accompanying coronal mass ejections (CMEs). If we could find any differences between them based solely on the light curves, we could hope to find similar differences on other stars. If not, that would hint at a universal flare profile.

To compile the sample, we use data from the \textit{Extreme ultraviolet Variability Experiment} \citep[EVE,][]{2012SoPh..275..115W}  instrument of the \textit{Solar Dynamics Observatory} (SDO).  EVE is an ultraviolet photometer with four different channels,  centered on 182,  256, 304, and  366\,\AA, measuring full disk irradiances with 0.25\,s cadence since 2010.  We use the 304\,\AA\, channel, which is centered on a He\,{\sc ii} line from the chromosphere.

\subsection{Flare extraction}

To locate the flares on the SDO/EVE light curves, we collected all the M and X class flares between 2010 and 2023 from the catalog of the \textit{Geostationary Operational Environment Satellite} (GOES)\footnote{\url{https://hesperia.gsfc.nasa.gov/goes/goes_event_listings/}}, resulting in 1136 events. Using this list, we downloaded Level 1 SDO/EVE time series from the LASP archive for the given days when flares occurred. Then, using the peak times from the GOES catalog, we extracted 9-hour cutouts, starting from 3 hours before the GOES peak. A 20-point (5\,second) running mean filter was applied to smooth the time series, and the negative flux values were removed.

The flare extraction from the SDO/EVE time series was carried out similarly to the TESS light curves. Using the nominal start and end times from the GOES catalog, we defined a quiescent region around the flare, and fitted it with a low-order polynomial, with the order determined by the BIC between 0 and 4. After subtracting this polynomial baseline, we fitted the flare template of \cite{davenport_template} to determine the \thalf time scale of the event. After shifting the flare peak time to zero, and scaling the amplitude to unity, we linearly interpolated each flare to a uniform time grid of 1000 points between $-2$ to 6\thalf, and removed a final linear trend fitted before -1 and after 5\thalf. 203 flares were discarded during the extraction (due to e.g., too many missing points, no dominant peak, failed baseline fit), resulting in a sample size of 933. We then visually inspected all of these flares, similar to the stellar case described in Sect.~\ref{sect:manual_vetting}, and removed all the incorrectly extracted ones. This resulted in a final sample size of 539. The main properties of these flares are summarized in Tab.~\ref{table:solar}, and the scaled flare shapes are available online.

The \thalf distribution of the SDO/EVE solar flares follows a log-normal distribution with $\mu = 0.88$ and $\sigma = 0.35$ in minutes, giving a median \thalf of 7.5\,minutes.

\begin{table*}
\caption{Parameters of the solar flares. The original SDO/EVE time series can be retrieved from \protect\url{https://lasp.colorado.edu/eve/data_access/eve_data/products/level1/esp/}, extended with the last column. The full table is available online.}
\label{table:solar}
\centering
\begin{tabular}{ccccc}
\hline \hline
GOES peak time & $t_{1/2}$ [min] & GOES class & CME & Time series URL\\
\hline
2010-05-05 17:19:00 & 3.3 & M1.2 & yes & \url{2010/esp_L1_2010125_007.fit.gz}\\
2010-06-12 00:57:00 & 3.5 & M2.0 & yes & \url{2010/esp_L1_2010163_007.fit.gz}\\
2010-08-07 18:24:00 & 36.5 & M1.0 & yes & \url{2010/esp_L1_2010219_007.fit.gz}\\
\dots & \dots & \dots & \dots & \dots\\
\hline
\end{tabular}
\end{table*}

\subsection{The effect of CMEs}
\label{sect:CME}
After collecting and scaling the solar flares from the 304\,\AA{} channel of SDO/EVE, we can look for differences in flare shape caused by different physical processes. One way to separate flares is whether they were accompanied by a CME.

\begin{figure*}[ht]
\centering
\includegraphics[width=1.95\columnwidth]{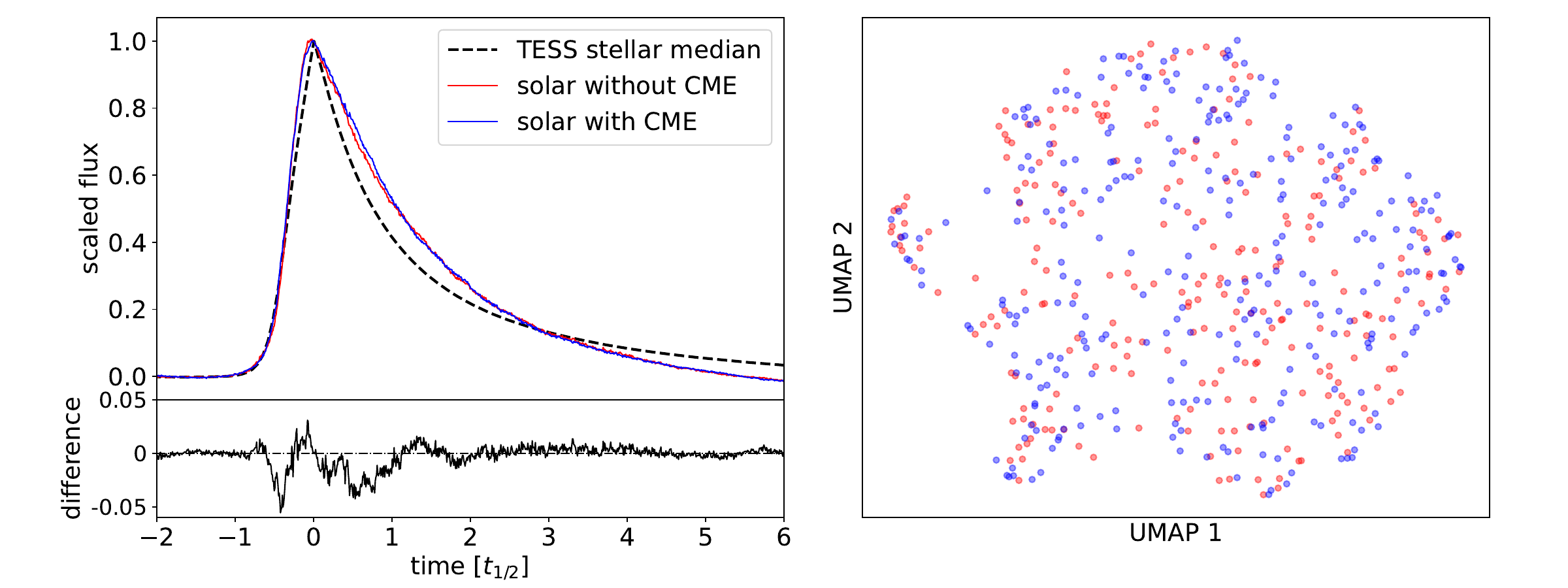}
\caption{Morphology of the solar flares observed in the 304\,\AA\, channel of SDO/EVE. \textit{Left:} Median shapes of solar flares with and without CMEs, and their difference. The median flare shape from TESS is also shown. \textit{Right:} UMAP projection of the scaled solar flare shapes. Blue and red points show events with and without CMEs, respectively.}
\label{fig:solar_shapes}
\end{figure*}

We test whether an accompanying CME influences the flare shape by contrasting the average shapes of flares with and without CMEs (so-called eruptive and confined flares in \citealt{2021ApJ...917L..29L}). We use the CME catalog of \cite{2009EM&P..104..295G}, created with data from the \textit{Solar and Heliospheric Observatory satellite} (SOHO), and flag the flares where a CME is reported in a $\pm2$\,hours interval near the flare peak time, and it is not labelled as a "(very) poor event" in the catalog. This results in 291 flares with CMEs and 248 flares without CMEs. The left panel of Fig.~\ref{fig:solar_shapes} shows the median flare shapes, with only a few percent difference between flares with and without CMEs. The average stellar flare shape from TESS is also shown, but due to the lower observing cadence and the different passband, it is hard to make a direct comparison. Following \cite{flaring_giants}, we quantify the difference between the average shapes of flares with and without CMEs using the sum of squared differences as a similarity metric. We calculate it for the median flare shapes, and compare it to a distribution of values from random shuffles of the dataset, mixing flares with and without CMEs together (see Sect.~3.2 of \citealt{flaring_giants}, and also their Appendix B). The resulting sum of squared differences is around the 70th percentile of the distribution from the random shuffles, indicating that the difference (if any) is weak, i.e. we get similar results from just randomly partitioning the flares into two groups. The right panel of Fig.~\ref{fig:solar_shapes} shows the UMAP projection of the 1000-dimensional dataset, and there is again no distinction between the two classes, in accordance with the previous result. If the two classes were noticeably different, they would separate more in the dimensionality-reduced space. The PCA representation of the dataset shows no distinction either.

Thus we failed to find any (simple) difference in the light curves of solar flares with and without accompanying CMEs. This hints that the diversity of stellar flares should probably not be attributed to CMEs, and that using high-resolution spectral time series remains the easiest way to reliably detect stellar CMEs \citep[see e.g.,][]{2019A&A...623A..49V, 2020MNRAS.493.4570L, 2022NatAs...6..241N}. A similar conclusion was reached by \cite{2016SoPh..291.1761H} for a sample of 42 X-class solar flares. They found that the only difference between flares with and without CMEs is coronal dimming in EUV. Coronal dimmings appear after flares with accompanying CMEs, they are mainly observable in hot coronal lines (e.g., Fe\,\textsc{xii} line at 193\,\AA, \citealt{2016SoPh..291.1761H}). In the stellar context, \cite{2021NatAs...5..697V} and \cite{2022ApJ...936..170L} presented observations of coronal dimmings on G--K--M stars, using X-ray and far-UV data. However, as coronal dimmings last for several hours -- an order of magnitude longer than the flaring time scale -- we cannot see them in the scaled flare shapes, as any variation after $\sim10$\thalf is removed with the baseline. However, we note that narrow-band EUV observations of the Sun are quite different from the white light time series we can study with TESS, so any implications should be interpreted with caution. Nevertheless, solar irradiation time series should be studied in greater detail, using data from different channels and instruments.

\section{Discussion}
\subsection{Physical background}

The evolution of a flare is usually divided into a heating and a cooling phase, separated by the peak time. In the heating phase, the heating rate exceeds the absorption rate, thus the temperature of the plasma rises. During the decay phase the heating rate drops below the sum of the conduction and radiative loss, and the plasma temperature drops.  The shape of the flare decay light curve is determined by the two cooling processes during the decay phase, i.e., radiative cooling and thermal conduction. In the initial part of the flare decay (roughly the first 20\% after peak time), the conductive cooling, while in the later part, the radiative losses play the main role typically resulting in a steep initial decay followed by a slower decline in flux
\citep{Aschwanden2004psci.book.....A}.  
The behavior of the flux during the decay phase depends on the ratio between cooling by radiation and cooling by thermal conduction, governed by the temperature and density of the emission region.
\cite{2021MNRAS.502.3922K} studied the decay phase of solar flares in several spectral bands -- in the 1600 and 304\,\AA{} channels that they used as Sun-as-a-star data, and in the 1700\,\AA{} channel, where the emission is associated with a similar temperature to that usually ascribed to M4-dwarf flares.
They found that the emission characteristics of various spectral bands are dependent on their formation temperatures and heights in the solar atmosphere. 
Namely, the decay rate during the first phase of cooling was slower for solar-like flares compared to M-dwarf flares, suggesting denser plasma in M-dwarf flares.
Furthermore, the study found differences in cooling behavior between solar flares and M-dwarf flares, with solar flares exhibiting more complex cooling patterns in the second phase.
This is exactly the behavior we can see in Fig.~\ref{fig:shapes_sequence} for the first time on other stars: hotter, solar-like stars typically show a slower decay in the conductive cooling phase and more complex cooling patterns compared to M-dwarfs suggesting that the plasma responsible for M-dwarf flare emission is denser than flares of hotter stars, and they potentially originate from a deeper layer of the stellar atmosphere.

Solar flares are seen as arcade-like structures, including multiple loops along the flare ribbon. \cite{2006ApJ...637..522W} showed that while single loop numerical models do not describe the X-ray light curve morphology and decay timescale of solar flares adequately, a multithread flare model works well. In these models, the decay time can be much longer than the single loop cooling timescale, as we see the superposition of multiple flaring loops. The quasi-periodic modulation of flares can also be described in the context of multithread models \citep[see e.g.,][]{2020ApJ...895...30R}. The change in the average shapes of stellar flares might also be related to multithread flares. In this context, the observed morphology can be linked to the physical structure of the flaring arcades, how many threads are formed and how frequently.

A possible source of flare shape variability may come from the different temperature evolution of flares on different stellar types, as suggested by \cite{2020ApJ...902..115H} on the basis of 44 flares observed simultaneously by TESS and Evryscope in $g'$ (29 of which are common with our sample). Their data suggest that the global color temperature of a flare does not depend on the stellar mass, while the color temperature at the flare peak increases with stellar mass, and thus with stellar \teff. If we approximate the flare spectrum with blackbody radiation, the largest fraction of flux in the TESS band is measured at around 5000\,K, and it declines for higher flare temperatures \citep[see Fig.~3 in][]{2020ApJ...902..115H}. Thus if the flare temperature changes, the fractional flux in the TESS band will change during the flare, resulting in systematic morphological differences in TESS observations. This shows the importance of time resolved, multiband observing campaigns of stellar flares \citep[see e.g.,][]{2013ApJS..207...15K, 2020ApJ...902..115H, 2024MNRAS.532.4436B, 2024MNRAS.529.4354J}.

\subsection{Possible future work}
In this work, we scaled and interpolated the extracted flares to make them comparable. This way the whole analysis was simplified, and we could use dimensionality reduction algorithms like PCA. However, this is not the only possible approach.

The analysis of one-dimensional shapes has a rich literature, albeit astronomical applications are rare (see, e.g., \citealt{2024arXiv240814466L}). It belongs to functional data analysis \citep{ramsay_fda}, a field of statistics that deals with curves and surfaces, where measurements are not isolated points, but continuous functions. One speciality of shape data is the invariance to certain transformations (e.g., translation, scaling). When comparing shapes, we should take these into account by selecting a suitable distance metric. In this work, we only used the simple Euclidean distance metric (L2 norm), which is the assumption behind many algorithms. This was only possible after prior scaling of the data. There are alternative metrics for functional data, e.g., dynamic time warping distance, cross-correlation distance, Procrustes distance. These are invariant to certain transformations, however, they are more computationally expensive to calculate. In functional data analysis, many classical methods have their counterparts, e.g., functional PCA, functional regression, $k$-Shape clustering \citep{10.1145/2949741.2949758}. This latter is an alternative to the popular $k$-Means clustering algorithm, but tailored to one-dimensional shapes. It was recently used by \cite{2023A&A...675A.130M} to cluster Stokes profiles from solar atmospheric simulations. One possible future avenue would be the application of functional data analysis techniques to the flare shape data, using different methods to represent or cluster the data.

One other way would be to keep the scaled profiles, and experiment with different dimensionality reduction algorithms. In this work, we applied PCA and UMAP. Another powerful method is the use of autoencoders \citep{1991AIChE..37..233K}. These are special neural networks, where the output to predict is the input itself. Autoencoders are comprised of an encoder part that transforms the input into a compact, low-dimensional latent space, and a decoder part, which transforms the data back. Once trained, the encoder can be used as a powerful dimensionality reduction algorithm. As the neural network can be arbitrarily complex, it can -- in theory -- learn complex representations, beyond the capability of PCA. Thus, more sophisticated dimensionality reductions algorithms might reveal more insights about the flare shape space. With a flexible algorithm, the scaling in duration and amplitude could also be omitted, making it possible to find more general correlations. \cite{2022MNRAS.509.5790L} studied the temporal morphology of radio pulses of the Vela Pulsar with similar goals. They worked with time series similar to flaring light curves, using variational autoencoders for dimensionality reduction, and self-organizing maps for the clustering of different pulse shapes. They succeeded in identifying different clusters and interpreted them as pulses originating from different heights in the pulsar magnetosphere.

Also, we did not devote much time to the analysis of interesting individual flaring objects. Apart from highly active stars, these also include hot or compact stars, where the tentative detection of flaring is intriguing in itself, however, its confirmation is more involved. It would require careful and detailed analysis, similar to the work of \cite{2024ApJS..271...57X} about flaring hot \mbox{subdwarfs} and white dwarfs with TESS.

\section{Summary}

In this study, we explored the information contained in the shapes of stellar flares. Our findings can be summarized as follows:
\begin{itemize}
    \item We searched for flares in the first five years (sectors 1--69) of the TESS mission, using 2-min cadence PDCSAP light curves.
    \item We used \flatwrm, a neural network-based algorithm to find flares. We re-trained \flatwrm specifically to TESS 2-min cadence data, by extending the previous training set with more than 4000 real TESS light curves, where we identified flares manually. We make this training set available online, which includes not only flaring stars, but also known astrophysical false positives.
    \item After filtering the \flatwrm flare candidates and manual vetting, we ended up with a high-purity catalog of $\sim 120,000$ flare events on $\sim 14,000$ stars (available as an online supplement). Besides basic parameters, we also extracted the scaled profile of each flare.
    \item  We found, that flare parameters -- equivalent duration, amplitude and \thalf{} --  correlate with \teff{} along the MS: with increasing temperature, equivalent duration and amplitude decrease, while \thalf{} increases.
    \item Flare shapes change with \teff{} as well -- flares of hotter stars are "fatter", wider for a few \thalf, but they decay more quickly, i.e., hotter, solar-like stars typically show a slower decay in the conductive cooling phase and more complex cooling patterns compared to M-dwarfs. This suggests that the plasma responsible for M-dwarf flare emission is denser than flares of hotter stars, and flares probably originate from a deeper layer of the stellar atmosphere.
    \item There was no indication of clustering in the flare shapes, the shapes seem to change gradually.
    \item The shapes of individual flares do not carry enough information to determine the physical parameters of their host.
    \item New flare templates were created for different \teff ranges.
    \item The PCA representation can be used to simulate realistic flare shapes, and to find flares similar to given input shapes.
    \item Using SDO/EVE data on solar flares, we analyzed the effect of coronal mass ejections (CMEs) on flare shapes and found no obvious difference between flares with/without CMEs, suggesting that the diversity of flares is not connected to CMEs.
\end{itemize}

As a future avenue, new observations from TESS, and also from the upcoming PLATO mission \citep{2014ExA....38..249R} will provide an ever growing catalog of stellar flares. Using a larger and more diverse sample, we can deepen our knowledge of magnetically active stars.

\section{Data availability}

The manually vetted flare catalog, the extracted flare shapes, the training set, and example Jupyter notebooks are available on the Zenodo service: \url{https://zenodo.org/records/14179313}

\begin{acknowledgements}
    We would like to thank the anonymous reviewer for the helpful comments, especially regarding the physical background of flare shapes. We thank G. Csörnyei for the helpful discussions and suggestions regarding data analysis methods.
    This research was funded by the Hungarian National Research, Development, and Innovation Office grants KKP-143986 and K-138962. Authors acknowledge the financial support of the Austrian--Hungarian Action Foundation grant 117öu4. K.V. is supported by the Bolyai János Research Scholarship of the Hungarian Academy of Sciences. B.S. was supported by the \'UNKP-22-3 New National Excellence Program of the Ministry for Culture and Innovation. On behalf of the “Looking for stellar CMEs on different wavelengths” project, we are grateful for the possibility of using the HUN-REN Cloud.
    Sz.S. acknowledges the support (grant No. C1791784) provided by the Ministry of Culture and Innovation of Hungary of the National Research, Development and Innovation Fund, financed under the KDP-2021 funding scheme.
      This work made extensive use of \texttt{numpy} \citep{numpy}, \texttt{scipy} \citep{scipy}, \texttt{pandas} \citep{mckinney-proc-scipy-2010}, \texttt{scikit-learn} \citep{scikit-learn}, \texttt{matplotlib} \citep{matplotlib} and \texttt{cmasher} \citep{2020JOSS....5.2004V}.
\end{acknowledgements}

\bibliography{bib}

\begin{appendix}
\section{Mock flare shape test}
\label{app:mock_flares}

To test what kind of variations we can recover from the flare shapes, we added artificial flares to 1000 real light curves with low, but non-zero flaring rates, then extracted them with the same procedure as the real events. The base flare shape model was the template of \cite{davenport_template}, with an additional temperature-dependent feature that contains a trend we try to recover. To create realistic datasets, we draw the \thalf and ED values of the base flare from a joint distribution (a Gaussian kernel density estimate of the real flare dataset), then calculate the $A$ amplitude analytically from the \cite{davenport_template} template as

\begin{equation}
    A = \frac{\mbox{ED}}{1.827 \cdot t_{1/2}}.
\end{equation}

\begin{figure}[hb]
\includegraphics[width=\columnwidth]{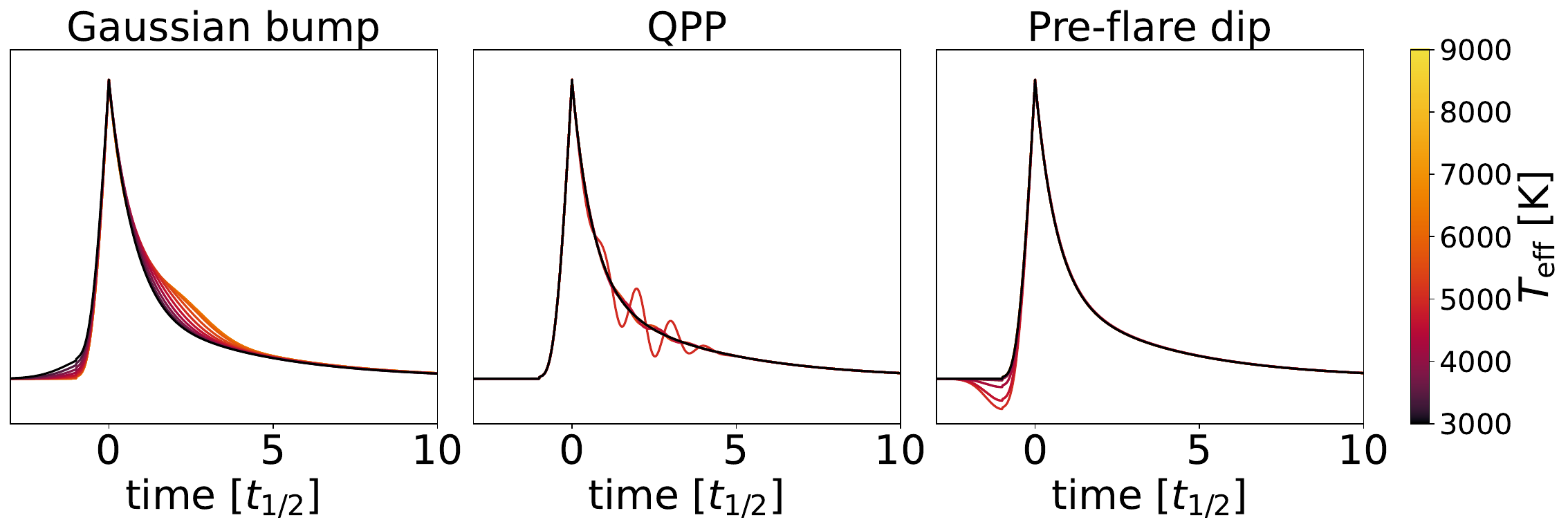}
\caption{The change of the injected flare shapes with effective temperature.}
\label{fig:mock_shapes}
\end{figure}

We inject 10 flares into each light curve at random times, taking care that they do not overlap with each other and real flare events. We consider three different artificial flare profiles, using the \teff of the injected star as a parameter. The purpose of this exercise is to recover the \teff using only the information artificially encoded in the flare profiles, illustrated in Fig.~\ref{fig:mock_shapes}.

\subsection{Gaussian bump}

The injected flare consists of a single peaked template $F(t_{\rm peak}, t_{1/2}, A)$ from \cite{davenport_template}, and an added Gaussian bump in the following form:

\begin{equation}
    F = F(t, t_{\rm peak}, t_{1/2}, A) + 0.1 \cdot A \cdot e^{-\frac{\left(t-t_{\rm bump}\right)^2}{2 t_{1/2}^2}},
\end{equation}
where
\begin{equation}
    t_{\rm bump} = \begin{cases} 0.0007 \cdot T_{\rm eff} - 2.2 & \mbox{if } T_{\rm eff} < 6000\,K \\ 2 & \mbox{if } T_{\rm eff} \ge 6000\,K. \end{cases}
\end{equation}
The bump has a width of \thalf, amplitude of 0.1 times the amplitude of the flare, and position determined by \teff. This dependence of the bump position on \teff is the trend hidden in the dataset that we try to recover.

\subsection{Quasi periodic pulsation (QPP)}

The injected flare is the sum of a single-peaked template $F(t_{\rm peak}, t_{1/2}, A)$ from \cite{davenport_template} and a localized sinusoidal in the following form:

\begin{multline}
    F = F(t, t_{\rm peak}, t_{1/2}, A) + \\
    A_{\rm QPP} \cdot \exp{\left[-\frac{\left(t - (t_{\rm peak} + 2 t_{1/2})\right)^2}{2 t_{1/2}^2}\right]} \cdot \sin{\left( \frac{2 \pi}{P_{\rm QPP}} t + \phi_{\rm QPP} \right)},
\end{multline}

where

\begin{equation}
    A_{\rm QPP} = 0.1 \cdot A \cdot \exp{\left[-\frac{\left(T_{\rm eff}-5000 K\right)^2}{2 (200 K)^2}\right]},
\end{equation}
and $P_{\rm QPP}$ is a period value drawn from a uniform distribution between 0.8 and 1.2 \thalf, and $\phi_{\rm QPP}$ is a phase value drawn from a uniform distribution between 0 and $2 \pi$. The \teff dependence lies in the amplitude of the pulsation (a sine curve with a Gaussian envelope), and peaks at 5000\,K. Note the random period and phase of the pulsation, which makes the problem more generic. QPPs are a matter of current research, see e.g., \cite{2018MNRAS.475.2842D} or \cite{2021SoPh..296..162R}.

\subsection{Pre-flare dip}
\label{app:preflaredip}
The most subtle alteration of the template of \cite{davenport_template}, with an added Gaussian dip before the rise phase:

\begin{equation}
    F = F(t, t_{\rm peak}, t_{1/2}, A) - A_{\rm dip} \cdot e^{-\frac{\left(t - (t_{\rm peak} - t_{1/2})\right)^2}{2 \left( 0.5 t_{1/2} \right)^2}},
\end{equation}

where

\begin{equation}
    A_{\rm dip} = 0.1 \cdot A \cdot e^{-\frac{\left(T_{\rm eff}-5000 K\right)^2}{2 (500 K)^2}}
\end{equation}

is the amplitude of the dip, which is the largest at $T_{\rm eff}=5000$\,K. Such dips have been observed by e.g., \cite{2014MNRAS.443..898L}.

\subsection{Recovery results}

\begin{figure}[hb]
\includegraphics[width=\columnwidth]{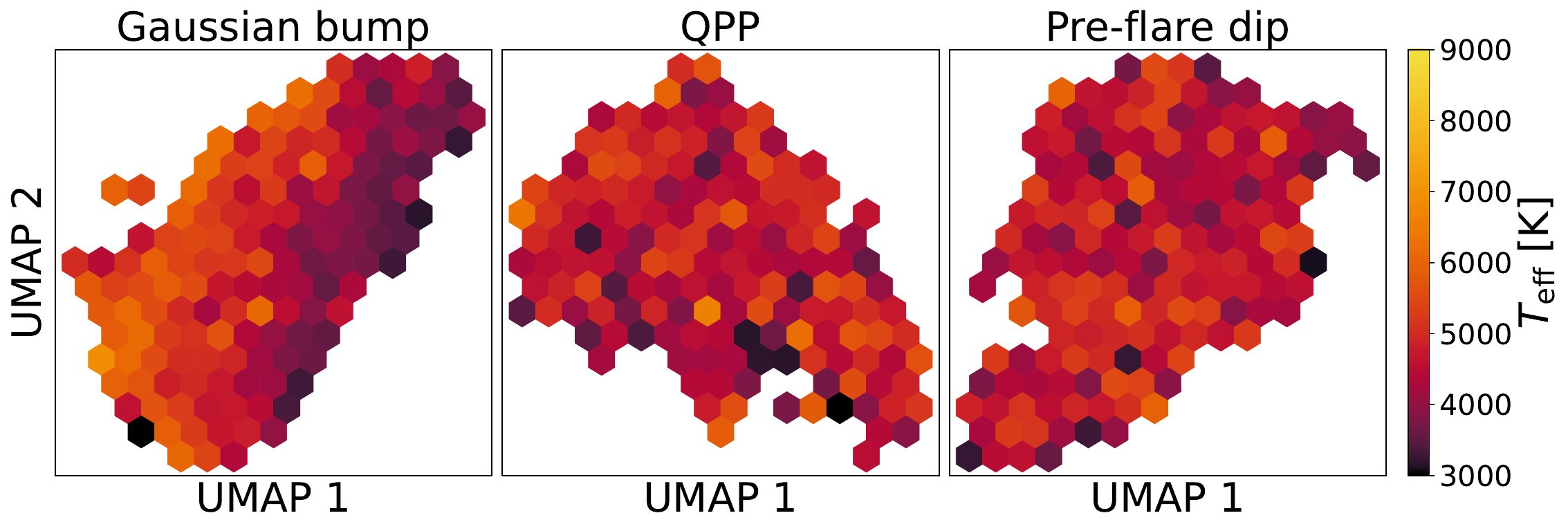}
\caption{UMAP projection of the recovered mock flares. Only the Gaussian bump shows a clear trend with \teff.}
\label{fig:mock_umap}
\end{figure}

To visualize the recovered temperature dependence, Fig.~\ref{fig:mock_umap} shows the two-dimensional UMAP projections of the extracted flare shapes. A clear \teff gradient would show that the hidden trends can be revealed by the dimensionality reduction technique. Adding the Gaussian bump created the only clearly noticeable effect. The appearance of these bumps resemble the trends seen in the case of real TESS flares in Fig.~\ref{fig:shapes_sequence}. The PCA projection shows similar trends to UMAP. The fact that we cannot recover the other two types of variability (QPP and pre-flare dip) indicates the limitation of the methods used in this study.

\end{appendix}

\end{document}